\documentclass[sigconf,10pt]{acmart}

\setcopyright{none}
\renewcommand\footnotetextcopyrightpermission[1]{}
\usepackage[left]{lineno}
\usepackage{gensymb}
\usepackage{multirow}
\usepackage{booktabs}
\usepackage{colortbl}
\usepackage{makecell}
\usepackage{enumitem}
\usepackage{xcolor}
\newcommand{\revcolor}{black}   
\newcommand{\revfont}{\sffamily}  
\newcommand{\revised}[2]{%
  {\begingroup
    \color{\revcolor}\revfont
    #2\ 
  \endgroup}%
}
\usepackage{tikz}
\usepackage{amsmath}
\usepackage[linesnumbered,ruled,vlined,algo2e]{algorithm2e}
 \usepackage[ruled, vlined, linesnumbered]{algorithm2e}
 \usepackage{algorithm}
\usepackage{algpseudocode}
\usepackage{multirow}
\usepackage{color}
\usepackage{filecontents}
\usepackage{subfig,graphicx}
\usepackage{xspace}

\newcommand\SystemName{\textsc{MetaGuardian}\xspace}

\newcommand\cparagraph[1]{\vspace{1.2mm}\noindent \textbf{#1.}}
\newcommand\sparagraph[1]{\vspace{1.2mm}\noindent \textbf{#1}}

\setcopyright{acmlicensed}
\copyrightyear{2018}
\acmYear{2018}
\acmDOI{XXXXXXX.XXXXXXX}

\acmConference[Conference acronym 'XX]{Make sure to enter the correct
  conference title from your rights confirmation emai}{June 03--05,
  2018}{Woodstock, NY}

\acmISBN{978-1-4503-XXXX-X/18/06}
\begin{document}
\def\allfiles{}
\date{}
\title{\SystemName: Enhancing Voice Assistant Security through Advanced Acoustic Metamaterials }
\author{Zhiyuan Ning}
\email{ningzhiyuan@stumail.nwu.edu.cn}
\orcid{0009-0009-4416-2080}
\affiliation{%
  \institution{NorthWest University}
  \city{Xi'an}
  \state{Shaanxi}
  \country{China}
}

\author{Zheng Wang}
\email{z.wang5@leeds.ac.uk}
\affiliation{%
  \institution{University of Leeds}
  \city{Leeds}
  \country{United Kingdom}}

\author{Zhanyong Tang}
\authornote{Corresponding author}
\email{zytang@nwu.edu.cn}
\affiliation{%
  \institution{NorthWest University}
  \city{Xi'an}
  \state{Shaanxi}
  \country{China}
}

\renewcommand{\shortauthors}{Ning et al.}
\begin{abstract}
We present \SystemName, a voice assistant (VA) protection system based on acoustic metamaterials. \SystemName can be directly integrated into the enclosures of various smart devices, effectively defending against inaudible, adversarial and laser attacks without relying on additional software support or altering the underlying hardware, ensuring usability. To achieve this, \SystemName leverages the mutual impedance effects between metamaterial units to extend the signal filtering range to 16-40~kHz to effectively block wide-band inaudible attacks. Additionally, it adopts a carefully designed coiled space structure to precisely interfere with adversarial attacks while ensuring the normal functioning of VAs. Furthermore, \SystemName offers a universal structural design, allowing itself to be flexibly adapted to various smart devices, striking a balance between portability and protection effectiveness. \revised{6}{In controled evaluation environments, \SystemName achieves a high defense success rate against various attack types, including adversarial, inaudible and laser attacks.}

\end{abstract}

\maketitle

\section{Introduction}
Voice assistants (VAs), such as Apple Siri, Google Assistant, and Amazon Alexa, have become widely integrated into mobile devices and smart home systems~\cite{r49,r50,r51,r52,r53,r54,r122}. However, their widespread adoption has also exposed them to various security threats~\cite{Backdoor,r8,r9}, including inaudible, adversarial, and laser-based attacks. Inaudible attacks\cite{longrange,r8,Backdoor,r9,nuit} embed malicious voice commands within ultrasonic or near-ultrasonic signals, making them imperceptible to human hearing but still recognizable by the VA\cite{commandersong,kenku,devil,qfa2sr}. Adversarial attacks involve carefully crafted audio inputs that sound benign to users but are intentionally designed to be misinterpreted as harmful commands by the VA. Laser-based attacks exploit amplitude-modulated light to remotely inject commands into the system. These attack methods are highly covert, making detection and mitigation particularly challenging.

Efforts have been made to create hardware- and software-based solutions to mitigate VA attacks. Software-based solutions focus on detecting attack signals entering the microphone and alerting the user to disable the voice assistant when a threat is identified. However, they face reliability issues caused by differences in microphone models and often struggle to effectively block attack signals while maintaining the normal functionality of the voice assistant~\cite{r5,r8,longrange,surfingattack}. Moreover, as mainstreamed Vas are usually closed systems, it is difficult to integrate and deploy a software-based solution. On the other hand, hardware-based defense solutions typically require modifications to commercial devices or the integration of additional active components. The former demands significant time and cost investments, while the latter may compromise the portability and practicality of mobile devices and likewise face reliability challenges in complex environments~\cite{r8,hardware1,hardware2,r5}.

\begin{figure}[t!]
    \centering
    \includegraphics[width=0.9\linewidth]{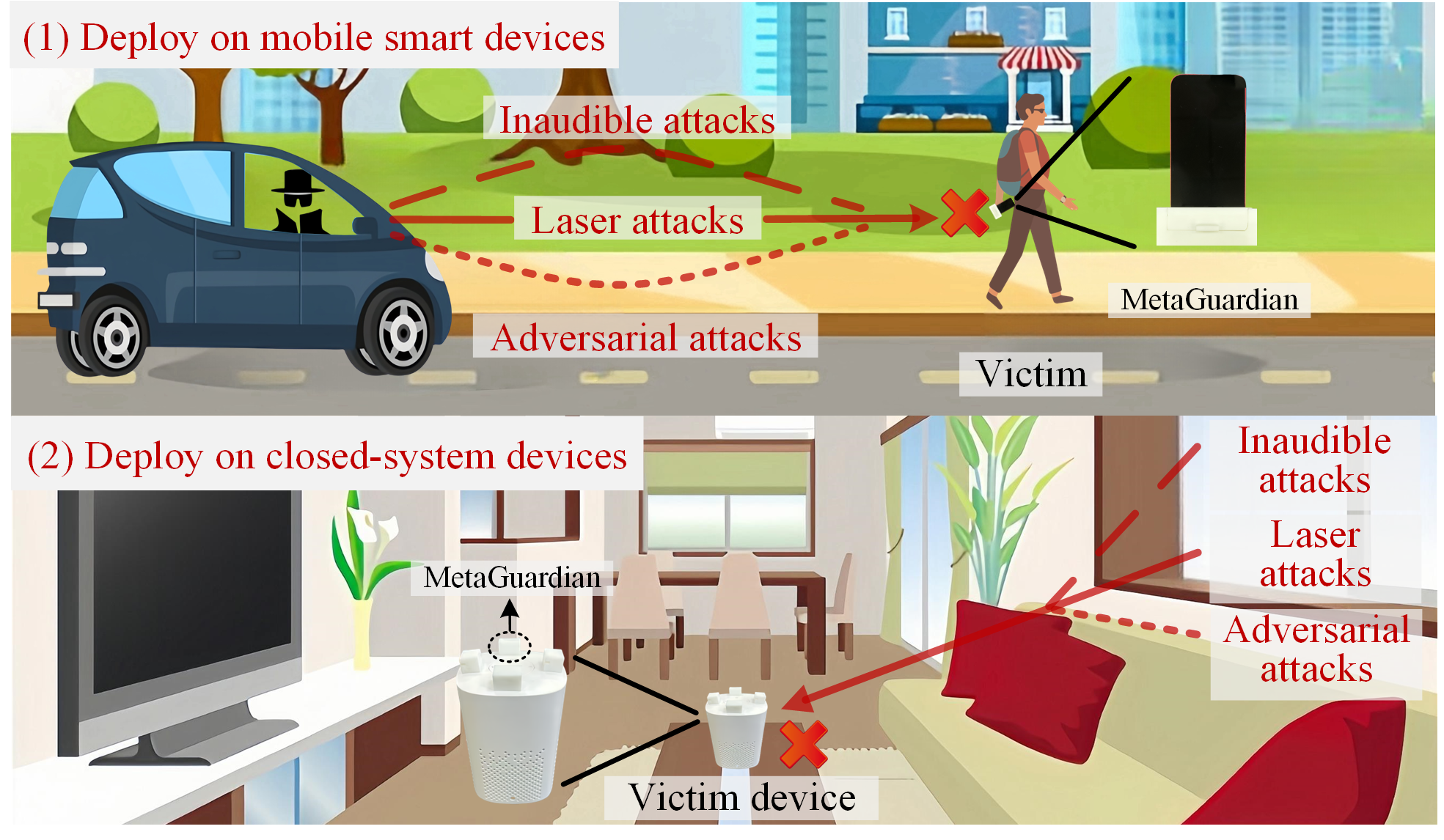}
\caption{Example deployment scenarios of \SystemName: (1) protecting mobile devices in public spaces without compromising portability and passivity; (2) interfering and blocking attacks on closed-system smart speakers.}
    \label{two possible scenario}
    \vspace{-10pt}
\end{figure}

Recent advancements in acoustic metamaterials~\cite{r17,r95,r57,r58,r59,r60,r61,r62,r63,r64} present a promising alternative to conventional defense strategies of VAs. Acoustic metamaterials manipulate sound waves through meticulously designed passive physical structures, enabling them to selectively block attack signals within specific frequency ranges while ensuring the normal operation of VAs, offering exceptionally high reliability. Unlike software-based solutions, acoustic metamaterials can effectively interfere with attack signals before they reach the microphone and do not rely on the device's operating system. This allows for broad deployment, even on closed-system devices. Their passive and compact nature enables seamless integration into smart device exteriors without requiring invasive modifications to the hardware, thereby preserving both circuit integrity and device portability. Figure \ref{two possible scenario} illustrates two typical defense scenarios that existing methods struggle to address effectively.

Although acoustic metamaterials offer the potential for protecting VAs from various attacks, developing a comprehensive defense system remains challenging. One major drawback of traditional acoustic metamaterials is their \textbf{narrow resonant frequency range}, which requires the combination of more than 13 units to effectively filter a broad spectrum of inaudible attack signals. This significantly compromises device portability. Additionally, when defending against \emph{adversarial attacks}, protective measures can interfere with the recognition accuracy of legitimate voice commands within overlapping frequency ranges, reducing usability in real-world applications. Furthermore, variations in device shape and microphone placement of the target device make it challenging to integrate acoustic metamaterials.

We present \SystemName, a new VA defense system based on acoustic metamaterials. \SystemName is designed to overcome the aforementioned limitations of traditional acoustic metamaterials. First, it implements a filtering mechanism leveraging the mutual impedance effect between metamaterial units. This enables effective filtering of inaudible attacks in the 16--40 kHz range with just \textbf{three units}, significantly reducing structural complexity while maintaining a compact volume of only $0.795 cm^3$.  Second, \SystemName enhances defense against \emph{adversarial attacks} by integrating labyrinth-style coiled acoustic metamaterial. This design selectively amplifies signals in the 2000--4000 Hz range, distorting critical frequencies to disrupt adversarial attacks while ensuring minimal impact on legitimate voice commands. Third, \SystemName provides a universal and portable design suitable for the two primary types of VA-equipped devices: mobile devices and smart speakers. The system adapts to structural and microphone placement characteristics, allowing for direct external mounting. Additionally, by strategically arranging metamaterials within reserved signal channels, the design ensures seamless transmission of legitimate commands while effectively disrupting attack signals.

We demonstrate that \SystemName can be manufactured using low-cost resin 3D printing technology and does not require users to train machine learning models. We evaluated \SystemName performance on nine smart devices against five adversarial attacks~\cite{kenku,alif,commandersong,devil,smack}, three inaudible attacks~\cite{Dolphinlang,nuit,lipread}, and one laser attack~\cite{lightcommands}. \revised{6}{In a controlled evaluation environment}, we show that \SystemName is compatible with various devices and can effectively defend against laser attacks. Moreover, within the effective attack distances identified by related studies, \SystemName successfully defends against all tested cases of five adversarial attacks and three inaudible attacks. Compared to existing defense solutions, \SystemName offers an innovative and efficient security protection mechanism for voice assistants.

This paper makes the following contributions:
\begin{itemize}[leftmargin=*]  
\item It presents the first acoustic metamaterial-based system that can effectively defend against inaudible, adversarial, and laser attacks without requiring software support or hardware modifications. 

\item It is the first to leverage mutual impedance effects to extend acoustic meta-matrial's filtering range to 16-40 kHz, blocking wide-band inaudible attacks while maintaining device functionality and portability.

\item It demonstrates how a portable VA protection system can be built through low-cost 3D printing.
\end{itemize}

\cparagraph{Online material} \revised{3}{The 3D printing CAD files for \SystemName and demonstration videos of system deployment can be downloaded from  \url{https://github.com/Meta-Guardian/MetaGuardian.}.}

\section{Background and Related Work} \label{chap:2}
In this section, we introduce the relevant background and compare \SystemName with prior defense strategies and alternative solutions. 


\subsection{Covert Attacks on Voice Assistants}  
Voice assistants (VAs) are vulnerable to three covert attack types: adversarial, inaudible, and laser. Unlike traditional transcription-based attacks, these can be executed without the victim's awareness, making them a greater threat~\cite{2024survey}.

\sparagraph{Adversarial attacks} embed malicious audio into conversations or music to deceive voice assistants into executing unintended commands~\cite{audios,Echo,transferable}. For example, CommanderSong~\cite{commandersong} hides adversarial perturbations in songs, while VRIFLE~\cite{vrifle} embeds them in user commands, enabling covert control.

\sparagraph{Inaudible attacks} \revised{1}{exploit ultrasonic frequencies, typically between 16 and 40 kHz, to deliver hidden voice commands. These attacks exploit weaknesses in how commercial microphones process sound, particularly in the early stages of the analog signal chain. In a typical microphone, an acoustic sensor such as a microelectromechanical systems diaphragm converts sound waves into electrical signals. These signals are then passed to a preamplifier. Ideally, the amplifier should increase the signal strength without altering its structure. However, due to limitations in device design, circuit implementation, and manufacturing processes, the amplifier often introduces nonlinear distortion when processing high-frequency signals. This distortion leads to the mixing of different frequencies. When an attacker sends an ultrasonic signal that carries a voice command, the nonlinear response of the amplifier causes frequency mixing. This process produces unintended low-frequency components that fall within the normal range of human speech. These components resemble the original voice command and are interpreted and executed by the voice assistant as if they had been spoken aloud by a person~\cite{Backdoor,r8,r9,lipread}.} \revised{1,6}{Although placing filters before the amplifier can help reduce the impact of inaudible attacks through analog signal processing, both modifying commercial microphones and using external filters have practical challenges. Modifying built-in microphones is difficult because they are usually integrated into closed proprietary chips that do not offer accessible interfaces for hardware changes. Furthermore, the wide variation in circuit designs across devices leads to high costs and poor adaptability. Using external filters also introduces complications, as these solutions require additional acoustic sensing components and separate power supplies. This increases system complexity and deployment costs, and makes them unsuitable for everyday use.
}



\sparagraph{Laser attacks} use modulated laser beams to inject commands into microphones, operating stealthily at distances over 100 meters, posing severe risks to privacy and device security~\cite{lightcommands}.

\subsection{Software-based Defenses}
Software-based approaches have been proposed to counter VA attacks. They employ varied tactics to counter voice threats. For examples, EarArray~\cite{eararray} detects inaudible attacks via signal timing differences across microphones. NormDetect~\cite{normaldetect} improves this by detecting missing features of the attack signal without heavy data needs. MVP-EARS~\cite{mvpear} reveals adversarial attacks through voice assistant transcription mismatches, and VSMask~\cite{vsmask} blocks them with real-time perturbations.

Software-based solutions often have limited reliability and may block attack signals at the cost of disrupting the normal operation of VAs. Their deployment is further challenged by the lack of access to internal systems on many commercial devices.
A key limitation is that detection methods based on signal features do not generalize well across different platforms, due to variations in microphone sensitivity and frequency response (see also Section~\ref{C1})~\cite{eararray,normaldetect}. As a result, these methods often fail in real-world settings. Some defenses try to stop inaudible attacks by disabling the VA entirely, which undermines normal usability~\cite{lipread,normaldetect,Dolphinlang}.
Moreover, as shown in Table~\ref{Common Smart Speakers' Audio Access Restrictions}, many commercial smart speakers restrict access to audio data for security reasons~\cite{normaldetect,eararray,robustdetection}. This restriction makes it difficult to test or deploy software defenses on real devices. Since simulation environments cannot fully reflect the diversity of hardware in actual products, evaluations based on them may lead to reduced effectiveness in practice.

\subsection{Hardware-based Defenses}
Hardware-based solutions \revised{7}{introduce changes to the hardware to defend against attacks on voice systems. For example, AIC~\cite{AIC} uses an additional speaker array to interfere with and block inaudible attacks. VocalPrint~\cite{vocalprint} uses millimeter wave probes to detect throat vibrations and confirm that the voice input is coming from a live human rather than a playback device. Similarly, the work presented in ~\cite{Sahidullah} uses a throat microphone to distinguish the user's voice from external speaker signals.}

As hardware-based defences require modifications to standard circuits or rely on non-portable active components, they have limited practical feasibility. Commercial devices usually adopt closed hardware architectures, making such invasive modifications challenging for end users. These modifications are often non-transferable across devices and can compromise functionality and stability, leading to compatibility issues~\cite{AIC,vocalprint}.
In addition, some hardware defenses depend on bulky, power-hungry components, such as speaker arrays or millimetre-wave radars~\cite{AIC,vocalprint}. These solutions hinder portability and restrict deployment, particularly in outdoor or mobile settings.
Furthermore, introducing additional hardware or circuit modifications increases system complexity and potential failure points. Attackers often exploit hardware-level traits, such as microphone non-linearity~\cite{Dolphinlang,lipread,Backdoor}. While these defenses can reduce certain risks, they may also create new vulnerabilities, such as instability, that could serve as new entry points for attacks. 

\begin{table}[t!]
    \scriptsize
    \caption{Smart speakers' audio access restrictions}
    \label{Common Smart Speakers' Audio Access Restrictions}
    \centering
    \setlength{\tabcolsep}{10pt} 
    \begin{tabular} {lllll}
    \toprule
    \textbf{Manuf.} & \textbf{Product Name} & \textbf{VA} & \textbf{Access Restr.}\\
    \midrule
    \rowcolor{gray!20}  Amazon & Echo Series & Alexa & No \\   
    Apple & HomePod Series &  Siri & No\\        
    \rowcolor{gray!20}Xiaomi & 	Xiaomi Speaker Series &  Xiao AI &  No\\    
     Huawei &Huawei AI Speaker Series  &  Xiaoyi &  No\\   
    \bottomrule
    \end{tabular}
\vspace{-10pt}
\end{table}

\subsection{Acoustic Metamaterials}
\SystemName is based on acoustic metamaterials and does not require changes to the software and hardware systems of the end-user devices. 
Using the macroscopic design of their internal structures, acoustic materials can modify the phase and amplitude of sound waves within particular frequency ranges~\cite{Remote,mitigating,hel1,hel2}, thus providing the potential to disrupt adversarial and inaudible attacks. Moreover, acoustic metamaterials can be constructed from opaque resin materials, which endows them with the ability to prevent the penetration of laser attacks. 

Unlike software and hardware defenses, acoustic metamaterials interact with sound waves purely through their passive physical structure, require no power supply, are compact in size, and can be placed externally to the device’s microphone, thereby circumventing the limitations of traditional solutions. However, they still face challenges such as expanding the filtering frequency range, maintaining device functionality, and achieving seamless integration across different devices, making it difficult to develop a comprehensive defense system.

\cparagraph{\revised{1}{Metamaterials vs. analog filters}}
\revised{1}{Acoustic metamaterials act like analog filters that can effectively block acoustic attacks. However, as discussed in Section 2.1, analog filters are difficult to deploy at scale in commercial devices. In contrast, acoustic metamaterials intercept attack signals before they reach the microphone, preventing effective attack components from being generated inside the device. Therefore, \SystemName require no modification to the microphone hardware, offering lower deployment costs and greater adaptability.}

\section{System Design for \SystemName} \label{chap:4}
\SystemName leverages acoustic metamaterials to build a VA defense system that is portable across devices and requires no modifications to the target device’s hardware or software. Developing \SystemName entails addressing three key challenges:
(1) \emph{Expanding the filter range} of acoustic metamaterial units to provide comprehensive protection against inaudible attacks;
(2) \emph{Achieving robustness} against adversarial attacks while preserving accurate recognition of legitimate audio;
(3) \emph{Ensuring portability} across diverse devices while balancing portability, functionality, and protection.
The following subsections (Sections~\ref{Challenge 1}–\ref{Challenge 3}) detail our solutions to these challenges.

\subsection{Expanding Filtering Range}  \label{Challenge 1}
Traditional acoustic metamaterials use Helmholtz-like resonators to filter ultrasound, but their narrow filtering bandwidth limits their ability to cover a wide range of inaudible attack frequencies. To address this limitation, we propose a solution based on the mutual impedance effect, which expands the bandwidth of the metamaterial units, enabling comprehensive defense against inaudible attacks.

\subsubsection{Principles and Narrowband Limitations of Metamaterials.} 
\begin{figure}[!t]
\centering
\subfloat[]{
		\includegraphics[scale=0.135]{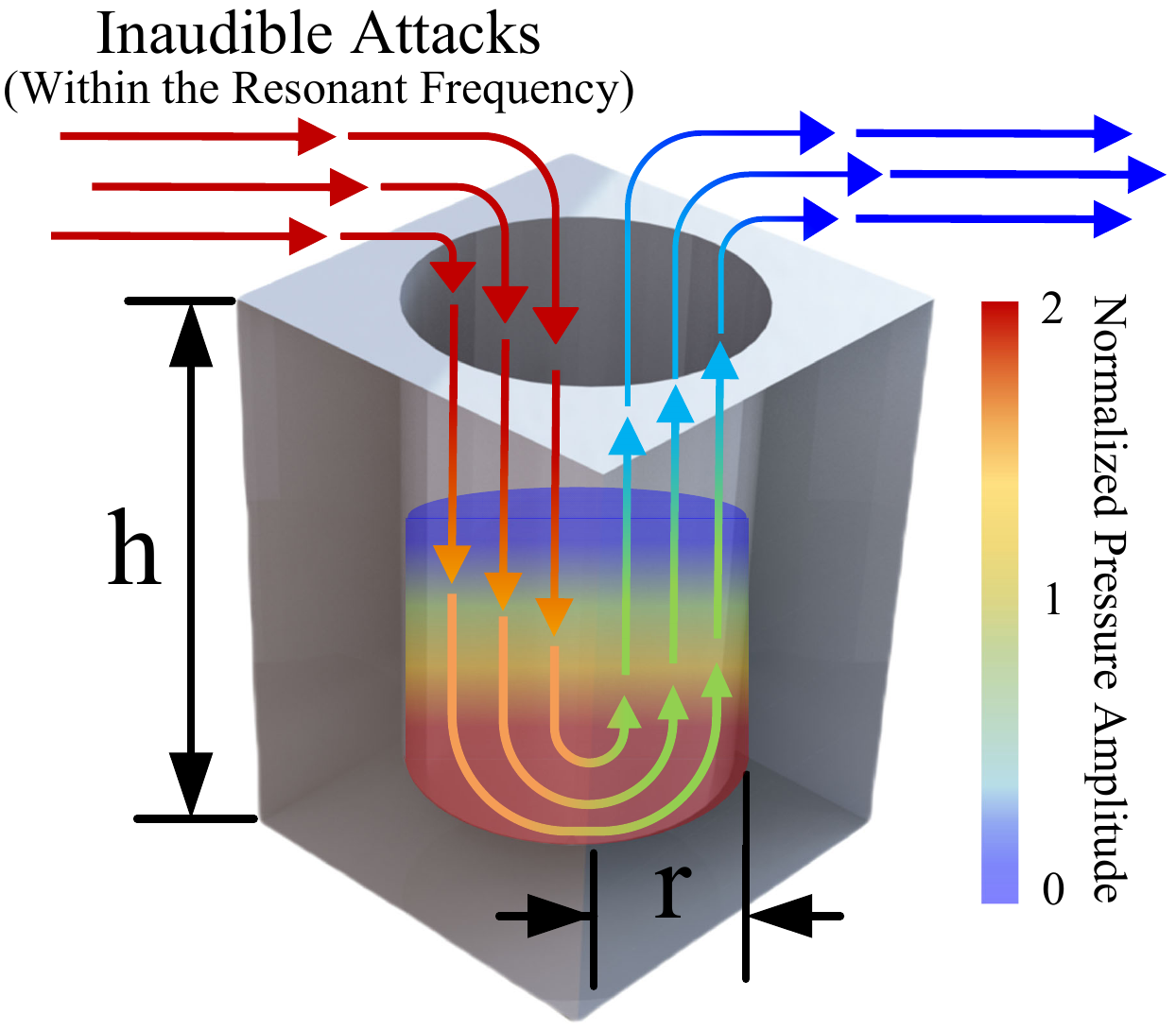}
  \label{meta1}}
\hfill
\subfloat[]{
		\includegraphics[scale=0.475]{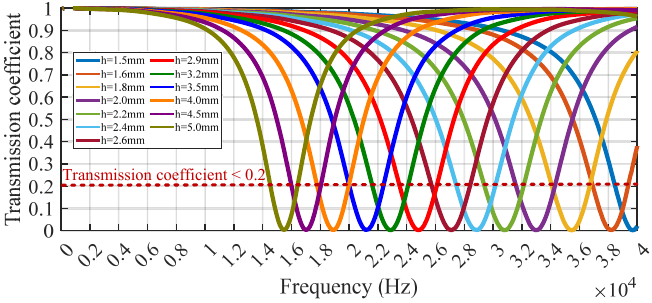}
  \label{meta1A}}
\caption{(a) Helmholtz-like acoustic metamaterial unit, (b) filtering effect of 13 units in 16-40~kHz.}
\label{Prin}
\vspace{-10pt}
\end{figure}

Acoustic metamaterials similar to Helmholtz resonators take advantage of their unique geometric structure to resonate with specific ultrasonic frequencies, allowing efficient filtering of ultrasound waves~\cite{mitigating}. As shown in Figure~\ref{meta1}, these acoustic metamaterials consist of a cylindrical cavity and a circular neck. When external sound waves enter the resonator, their energy interacts with the air inside the cavity, leading to significant absorption of sound wave energy near the resonant frequency and greatly attenuating the energy of external sound waves passing through the cavity.

Therefore, to match the resonance frequency of ultrasound, it is necessary to design an appropriate cavity structure of the acoustic metamaterial. \revised{7}{The resonance frequency \(f_0\) of the acoustic metamaterial can be calculated using the following formula~\cite{r24}:}  
\begin{equation}\label{eqn-1}  
f_0 = \frac{v}{4(h + r)}  
\end{equation}  
where \(v\) is the speed of sound in air (typically 343~m/s), \(h\) is the depth of the cylindrical cavity, and \(r\) is the radius of the narrow neck. By adjusting these parameters, highly efficient filters targeting specific ultrasonic frequency bands can be precisely designed.  

However, the resonance frequency of acoustic metamaterials is highly dependent on the precise matching of their geometric structure, which limits a single fixed-structure metamaterial unit to filtering a relatively narrow frequency range. Although LLOYD et al.~\cite{mitigating} pointed out that when \(r = 1.5 \, \text{mm}\), acoustic metamaterials can achieve a wider filtering bandwidth, the resonance of a single metamaterial unit is still confined to a frequency range of 1-2 kHz. In contrast, the frequency range of inaudible attacks spans a much broader range of 16-40~kHz. As shown in Figure~\ref{meta1A}, to reduce the transmission coefficient to approximately 0.2 and effectively defend against such a wide range of inaudible attacks~\cite{mitigating}, around 13 metamaterial units are required. This significantly increases system complexity and dramatically reduces portability, making it difficult to integrate with devices.

\subsubsection{Broadband Filtering via Mutual Impedance Effect} 

Recent studies~\cite{mitigating,Mutualimpedanceeffects} have shown that mutual impedance can tune the resonant frequency and broaden the frequency range. Inspired by this, we leverage this effect to achieve broadband filtering with fewer units, breaking the limitations of traditional methods.

Specifically, mutual impedance enhances the total impedance \( Z_{\text{total}} \) of the system, which influences the resonant frequency through the following equation: \(
f_r = \frac{1}{2\pi} \sqrt{\frac{1}{m_{\text{eff}} Z_{\text{total}}}}
\), \revised{7}{where $m_\text{eff}$ represents the effective mass, reflecting the inertia of the structure under a specific vibration mode. It is determined jointly by the material and structure of the metamaterial unit.}
As \( Z_{\text{total}} \) increases, the resonant frequency extends toward the lower frequency range, thereby expanding the overall frequency range. The total impedance \( Z_{\text{total}} \) of the system can be expressed as:\(
   Z_{\text{total}} = \sum_{i=1}^{N} Z_i + Z_{\text{mutual}} 
\). This indicates that enhancing the mutual impedance effect is essential for expanding the resonant frequency range.

To achieve this, we further investigated the correlation between the strength of the mutual impedance effect and the spatial configuration of metamaterial units, leading to a critical finding: \textit{The separation and arrangement of the units significantly impact the mutual impedance effect.}

\cparagraph{The effect of unit spacing on mutual impedance} We have ascertained that the magnitude of the mutual impedance effect is intricately associated with the distance \( S \) that separates the units. Specifically, the mutual impedance exhibits an inverse proportionality to this distance. This correlation is articulated by the following equation: \begin{equation}\label{distence}
Z_{\text{mutual}} \propto \frac{1}{S}
\end{equation} wherein \( S \) signifies the distance between adjacent units. We posit that the fundamental cause of this correlation is attributable to the influence of \( S \) on the coupling effect between units. Reduced separation between units intensifies the coupling effect, thereby augmenting the impedance interaction and the resultant mutual impedance. 

To further substantiate this relationship, we constructed a spatial correlation model based on acoustic coupling theory. Analogous to the mutual inductance theory in electromagnetics, the acoustic mutual impedance can be represented as the spatial integral of the sound pressure fields of adjacent units: \begin{equation}\label{electromagnetics}
Z_{\text{mutual}} = \frac{1}{j\omega} \int_{V} \left( p_1 \cdot v_2^* \right) dV
\end{equation} Here, \( p_1 \) represents the sound pressure radiated by the first unit, \( v_2^* \) is the complex conjugate of the vibration velocity of the adjacent unit, and \( \omega \) is the angular frequency. When the unit spacing is much smaller than the wavelength of the sound wave (\( S \ll \lambda \)), the sound pressure field approximately follows a spherical wave decay (\( p \propto 1/S \)), meaning that the closer the distance, the stronger the sound pressure. Additionally, the vibration velocity is in phase with the sound pressure, indicating that an increase in sound pressure simultaneously enhances the vibration velocity. Therefore, the integration result satisfies Equation~\ref{distence}, confirming the inverse relationship between mutual impedance and distance.

In addition, we developed an equivalent RLC circuit model, treating each metamaterial unit as a resonant RLC circuit and using mutual inductance \( M \) to represent the mutual impedance: \begin{equation}\label{RLC}
Z_{\text{mutual}} = j\omega M = j\omega \left( k \sqrt{L_1 L_2} \right)
\end{equation} Here, the coupling coefficient \( k \propto 1/S \), which is also inversely related to the distance. This model further provides theoretical support for the regulation of mutual impedance.

To optimize the mutual impedance effect, we set \( S \) to \textbf{0.1 mm}, a distance that maximizes the mutual impedance while meeting the precision requirements of 3D printing, ensuring that the units do not overlap and avoiding structural interference. 

\cparagraph{The effect of unit arrangement on mutual impedance} The arrangement of the units also plays a key role in the strength of the mutual impedance effect. Through an analysis of different arrangements, we found that a linear arrangement significantly enhances the mutual impedance effect. This is because a linear arrangement only involves direct coupling between adjacent units, thereby avoiding the weakening of the mutual impedance effect caused by complex interactions among multiple units in more intricate layouts, such as circular arrangements. The mutual impedance under different configurations can be expressed as follows:   
\begin{equation}\label{Linear}
Z_{\text{mutual}} =
\begin{cases} 
\frac{1}{S} \cdot \alpha, & \text{Linear} \\[8pt]
\frac{1}{S} \left( \alpha \cdot f_{\text{loss}}(N) \right), & \text{Circular}
\end{cases}
\end{equation} 
Where \( \alpha \) is the coupling coefficient, which depends on the unit arrangement structure, material properties, and the surrounding medium of the units. In a linear arrangement, \( \alpha \) is primarily determined by direct coupling between adjacent units. For a circular arrangement, there exists multi-path interference between units, and the mutual impedance weakening factor \( f_{\text{loss}}(N) \) can be expressed as: \begin{equation}\label{loss}
f_{\text{loss}}(N) = \frac{1}{N} \sum_{m=1}^{N} \sin^2\left(\frac{\pi m}{N}\right)
\end{equation} 

As the number of metamaterial units $N$ increases, the function $f_{\text{loss}}(N)$ decreases, indicating that phase mismatches between non-adjacent units cause energy loss, weakening the mutual impedance effect in circular arrangements. Moreover, $f_{\text{loss}}(N)$ satisfies $0 < f_{\text{loss}}(N) < 1$, showing that additional coupling in circular arrangements reduces overall mutual impedance. To verify this, we tested filtering effects for different arrangements and spacings in experiments (see Section~\ref{A1}).

\begin{figure}[!t]
\centering
\subfloat[]{
		\includegraphics[scale=0.44]{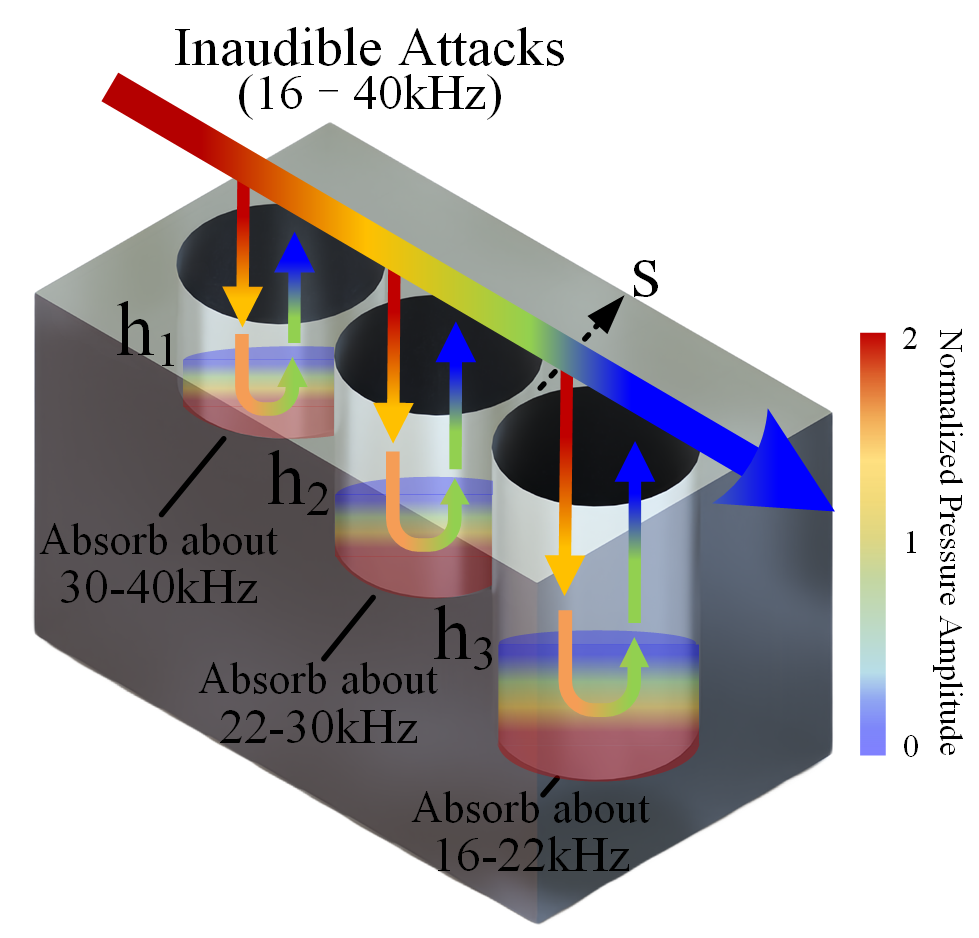}
  \label{meta2}}
\hfill
\subfloat[]{
		\includegraphics[scale=0.48]{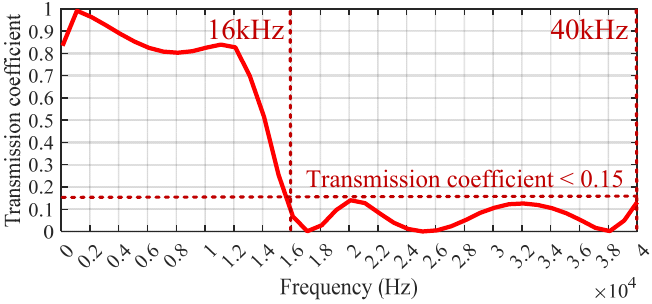}
  \label{meta2A}}
\caption{(a) Inaudible Attack Defense Metamaterial(IADM), (b) its filtering effect in 16-40~kHz range.}
\label{IADM}
\vspace{-10pt}
\end{figure}

By linearly arranging metamaterial units with a spacing of 0.1 mm, COMSOL simulations show that the resonance frequency range of a single unit expands nearly fourfold. Three units of different heights ($h_1 = 2mm$, $h_2 = 3.2mm$, $h_3 = 4.8mm$) were selected to cover the inaudible attack frequency band of 16–40 kHz, forming the Inaudible Attack Defense Metamaterial (IADM). As shown in Figure \ref{meta2A}, simulation results indicate that the IADM reduces the ultrasonic wave transmission coefficient in this band to below 15\%, demonstrating significant defense effectiveness.

\subsection{Achieving Robustness} \label{Challenge 2}
To address adversarial attacks, we propose a coiling-up space-structured metamaterial capable of amplifying signal amplitude within a specific frequency range, thereby disrupting or weakening the critical features of attack signals and neutralizing adversarial attacks~\cite{devil,sirenattack}. However, if the interference frequency range is crucial for legitimate audio, it may affect the normal operation of the voice assistant. Therefore, precise analysis and the design of metamaterials tailored to that frequency range are necessary.

\subsubsection{Selection of Interference Frequency Bands}
To ensure the intelligibility of legitimate audio while effectively interfering with adversarial attack signals, it is crucial to select an appropriate interference frequency band. The clarity of human speech (100-4000 Hz) primarily depends on the first (F1: 100-1000 Hz) and second formants (F2: 1000-2000 Hz) \cite{format1,format2}, while the 2000-4000 Hz range mainly carries consonant details, contributing only about 10\% of the total speech information entropy (\( H_{\text{high}} / H_{\text{total}} \approx 10\% \)) \cite{importantHz1,importantHz2}. \revised{2}{Conversely, adversarial attacks typically embed perturbations in the 2000-4000 Hz frequency range to enhance their stealth, allowing them to interfere with the normal operation of speech recognition systems without being easily perceived by the human ear~\cite{Learning,Towardrobust,Robustaudio}.} Therefore, interfering within this frequency range can maximize the suppression of adversarial attacks while preserving essential speech content.  

Coiling-up space-structured metamaterials can effectively neutralize adversarial attack signals by amplifying perturbations and introducing nonlinear distortion. Adversarial attacks typically add a small perturbation \( \delta(t) \) to the legitimate audio, with its power significantly lower than the original signal: \begin{equation}\label{adv}
x_{\text{adv}}(t) = x_{\text{clean}}(t) + \delta(t), \quad P_{\delta}(f) \ll P_{x_{\text{clean}}}(f).
\end{equation}

Metamaterials utilize frequency-selective resonance to significantly amplify signals within a specific band. Given a transmission gain \( H(f) \), the processed signal is expressed as: \(x_{\text{meta}}(t) = \mathcal{F}^{-1} \left\{ H(f) X_{\text{adv}}(f) \right\}.\), when \( H(f) \gg 1 \) (applied only to the 2000-4000 Hz range), the perturbation \( \delta(t) \) is greatly amplified, introducing nonlinear distortion that disrupts attack features: \begin{equation}\label{disturb}
\tilde{\delta}(t) = \mathcal{F}^{-1} \left\{ H(f) \Delta(f) \right\}.
\end{equation}

Therefore, this metamaterial design effectively weakens adversarial attacks.

\cparagraph{\revised{2}{Advantages over direct filtering}}
\revised{2}{While modifying the IADM structure can also filter out the frequency band used in adversarial attacks, this band also carries important information for automatic speech recognition and speaker identification. As a result, direct filtering is likely to degrade these functions and significantly impair daily usage. In contrast, the space-wrapping metamaterial used by \SystemName selectively interferes with critical features of attack signals. Although it may introduce some impact on speech, it preserves legitimate audio to the greatest extent, making it a more practical and effective defense against adversarial attacks. In Section~\ref{A2}, we provide empirical evidence showing the advantage of \SystemName over direct filtering.}

\begin{figure}[!t]
\centering
\subfloat[]{
		\includegraphics[scale=0.110]{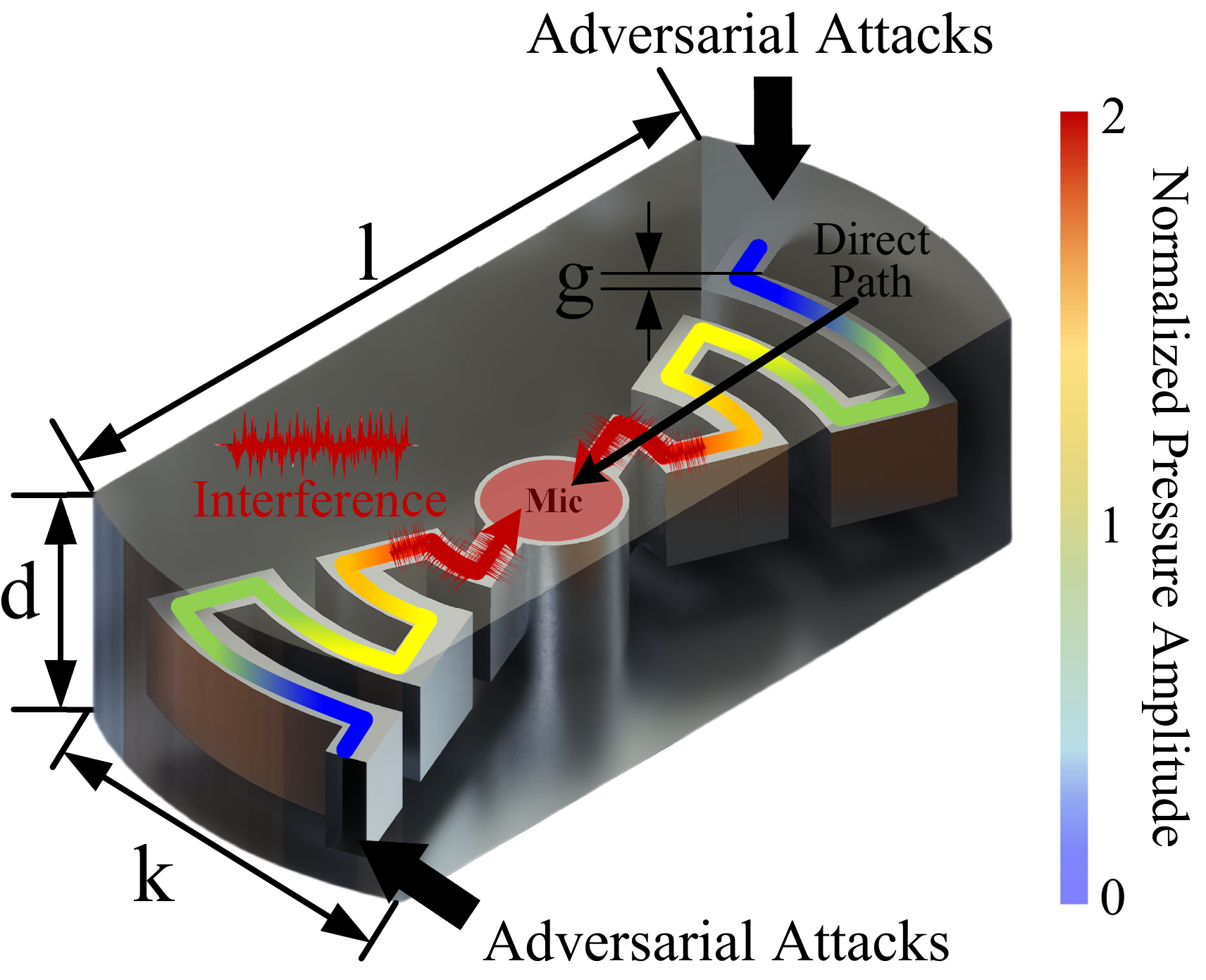}
  \label{meta3A}}
\hfill
\subfloat[]{
		\includegraphics[scale=0.505]{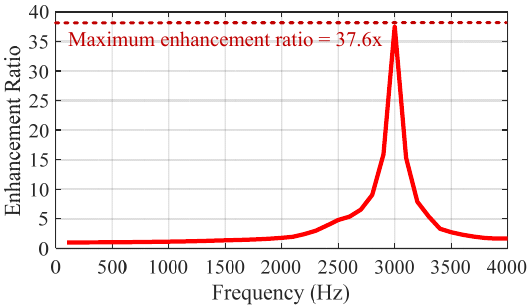}
  \label{meta3B}}
\caption{(a) Adversarial Attack Defense Metamaterial(AADM), (b) its interference effects.}
\label{AADM}
\vspace{-10pt}
\end{figure}

\subsubsection{Metamaterial Design for Adversarial Attack Defense}  
We propose a novel coil space-structured acoustic metamaterial to enhance audio signals in the 2000-4000 Hz frequency range and achieve interference effects. As shown in Fig.~\ref{meta3A}, the metamaterial adopts a slender design, effectively reducing its size and improving portability. It consists of two sets of helical spatial structures that extend the propagation path of sound waves to regulate the resonance frequency, generating strong resonance within the target frequency band. During resonance, the acoustic energy is concentrated and amplified, thereby enhancing signals in this frequency range to interfere with adversarial signals. The dimensions of the metamaterial are as follows: length \(l = 15 \, \text{mm}\), width \(k = 7.65 \, \text{mm}\), height \(d = 4.75 \, \text{mm}\), and internal channel width \(g = 0.8 \, \text{mm}\).

Initially, the resonant frequency \(f_r\) of the acoustic metamaterial determines its response and amplification capability for specific frequency signals, and is closely related to the internal path length \(L_{\text{coiled}}\). The formula for calculating the resonant frequency \(f_r\) is: \begin{equation}\label{coiled}
f_r = \frac{c}{4L_{\text{coiled}}}
\end{equation} where \(c\) denotes the speed of sound in air, which is 343 m/s, and \(L_{\text{coiled}}\) represents the length of the coiling path within the metamaterial. As the frequency of the sound wave approximates the resonant frequency, the metamaterial demonstrates its most potent energy response, thereby amplifying signals within that particular frequency spectrum. By judiciously selecting an appropriate path length \(L_{\text{coiled}}\), the resonant frequency of the metamaterial can be modulated to align with the designated frequency range.

In the proposed design, the specified target frequency range is 2000-4000~Hz, thereby setting the resonant center frequency as noted in \(f_r = 3000\, \text{Hz}\). Using Equation \ref{coiled}, the calculated coil path length is determined to be as indicated in \(L_{\text{coiled}} = 28.5 \, \text{mm}\). This configuration ensures that the metamaterial produces a substantial enhancement effect within the designated target frequency range.
Subsequently, after determining \(L_{\text{coiled}}\), the sound pressure amplification factor \(G\) is calculated using the following equation: \begin{equation}\label{eqn-5}
G = \frac{n_r}{\lambda_0} \cdot \sqrt{\frac{2 \rho c^2}{\lambda_0^2}}
\end{equation} In this context, the refractive index \(n_r = \frac{L_{\text{coiled}}}{L_{\text{blue}}}\) is defined as the quotient of the propagation speed of sound waves within the metamaterial and their speed in air. By calculating the path length ratio shown in Figure~\ref{meta3A}, this refractive index can be estimated. When an adversarial attack passes through the metamaterial with a high refractive index \(n_r\), the sound pressure is excessively amplified, leading to distortion. This metamaterial is designated as the \textit{Adversarial Attack Defense Metamaterial} (AADM). 
 
The COMSOL simulation results (Figure~\ref{meta3B}) are consistent with theoretical predictions, showing enhanced sound energy within the 2000-4000 Hz frequency range, with a maximum gain of 37.6 times at 3000 Hz. Subsequently, we also verified in Section~\ref{A2} and Section~\ref{B1} that AADM effectively defends against adversarial attacks while maintaining the integrity of legitimate audio signals.

\begin{figure}[!t]

\centering
\subfloat[]{
\includegraphics[scale=0.082]{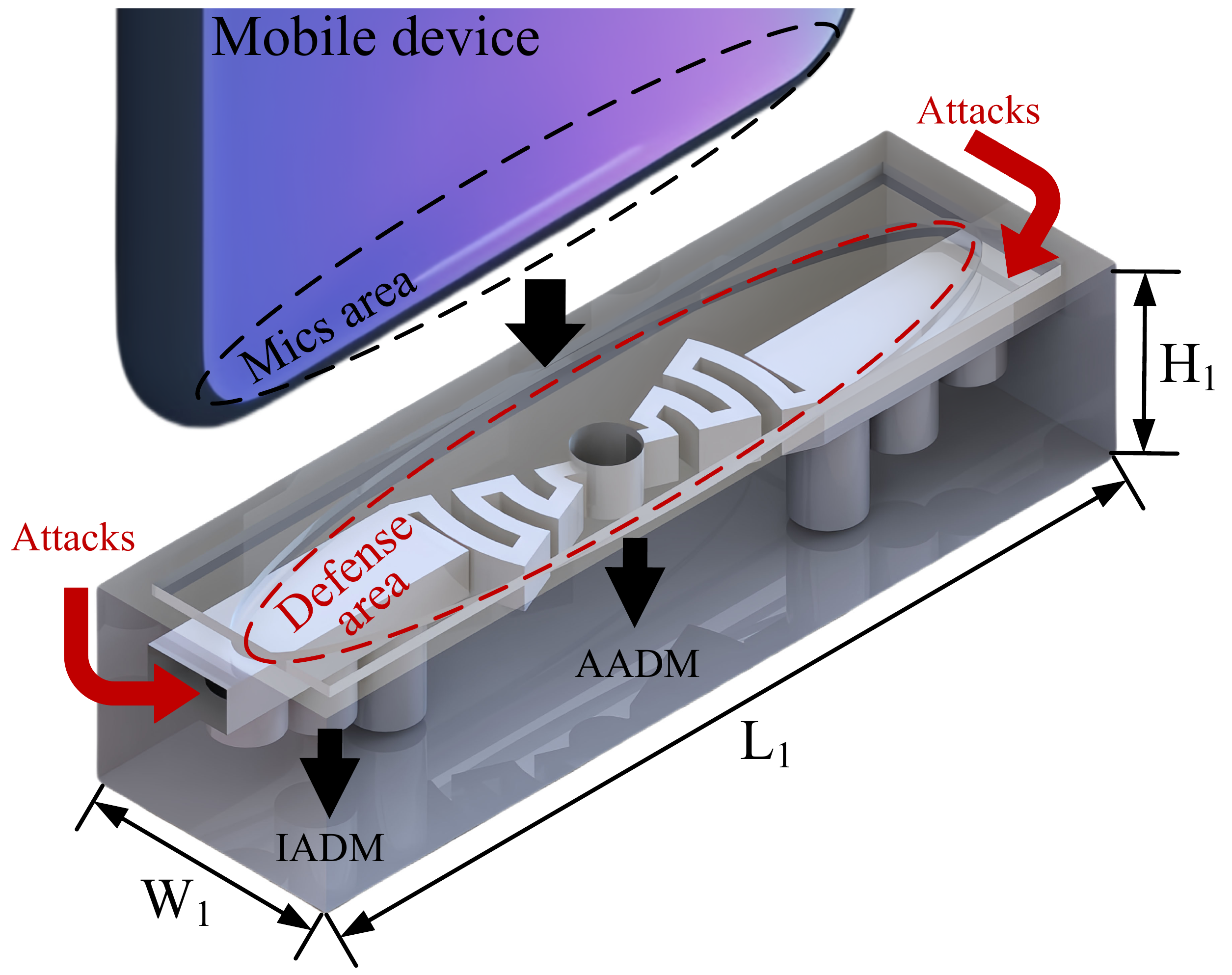}
  \label{meta44}}
\hfill
\subfloat[]{
		\includegraphics[scale=0.118]{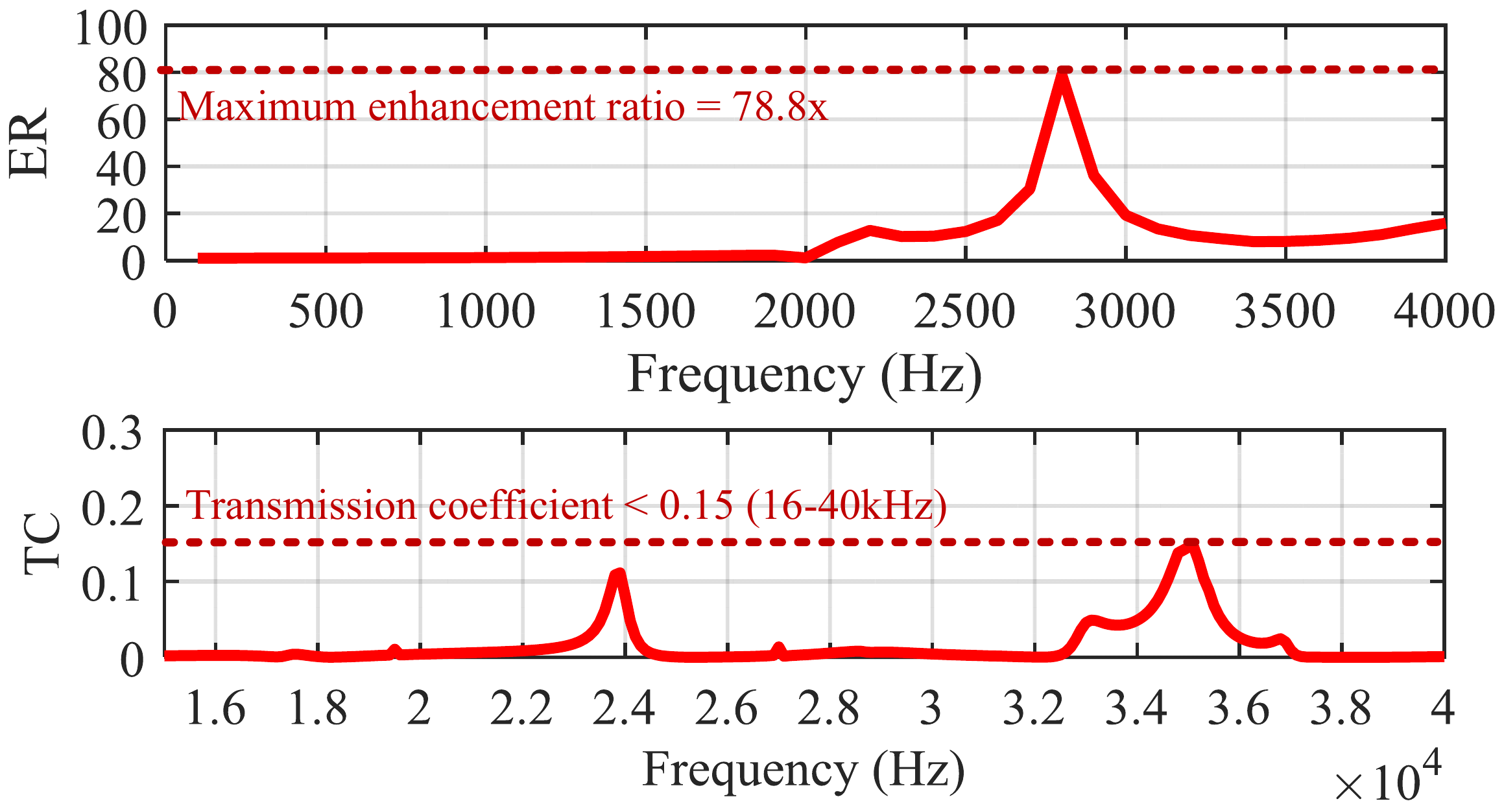}
  \label{meta44A}}
\caption{(a) Mobile devices structure design, (b) its filtering and interference effects.}
\label{mobile}
 \vspace{-10pt}
\end{figure}

\subsection{Ensuring Portability}\label{Challenge 3}

Although IADM and AADM each perform well in defense, \SystemName faces practical challenges due to significant differences in structure and microphone layouts among mainstream voice assistant devices. The key issue is how to integrate both into a universal defense structure that balances effective protection with device portability and functionality. To address this, we analyzed the structural features of mobile devices and smart speakers and designed dedicated universal defense solutions for each.

\begin{figure}[!t]

\centering
\subfloat[]{
\includegraphics[scale=0.0815]{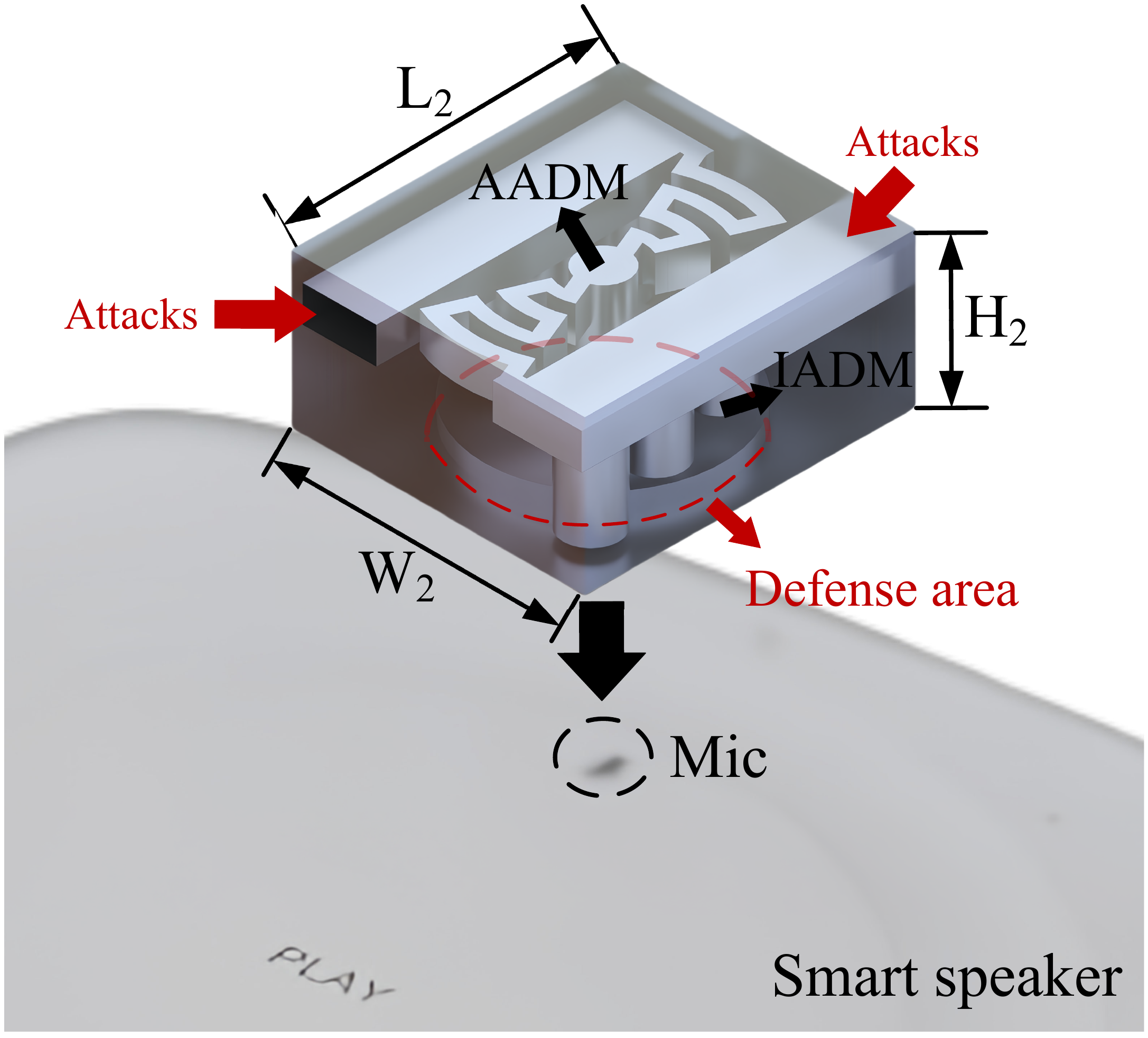}
  \label{meta5}}
\hfill
\subfloat[]{
		\includegraphics[scale=0.12]{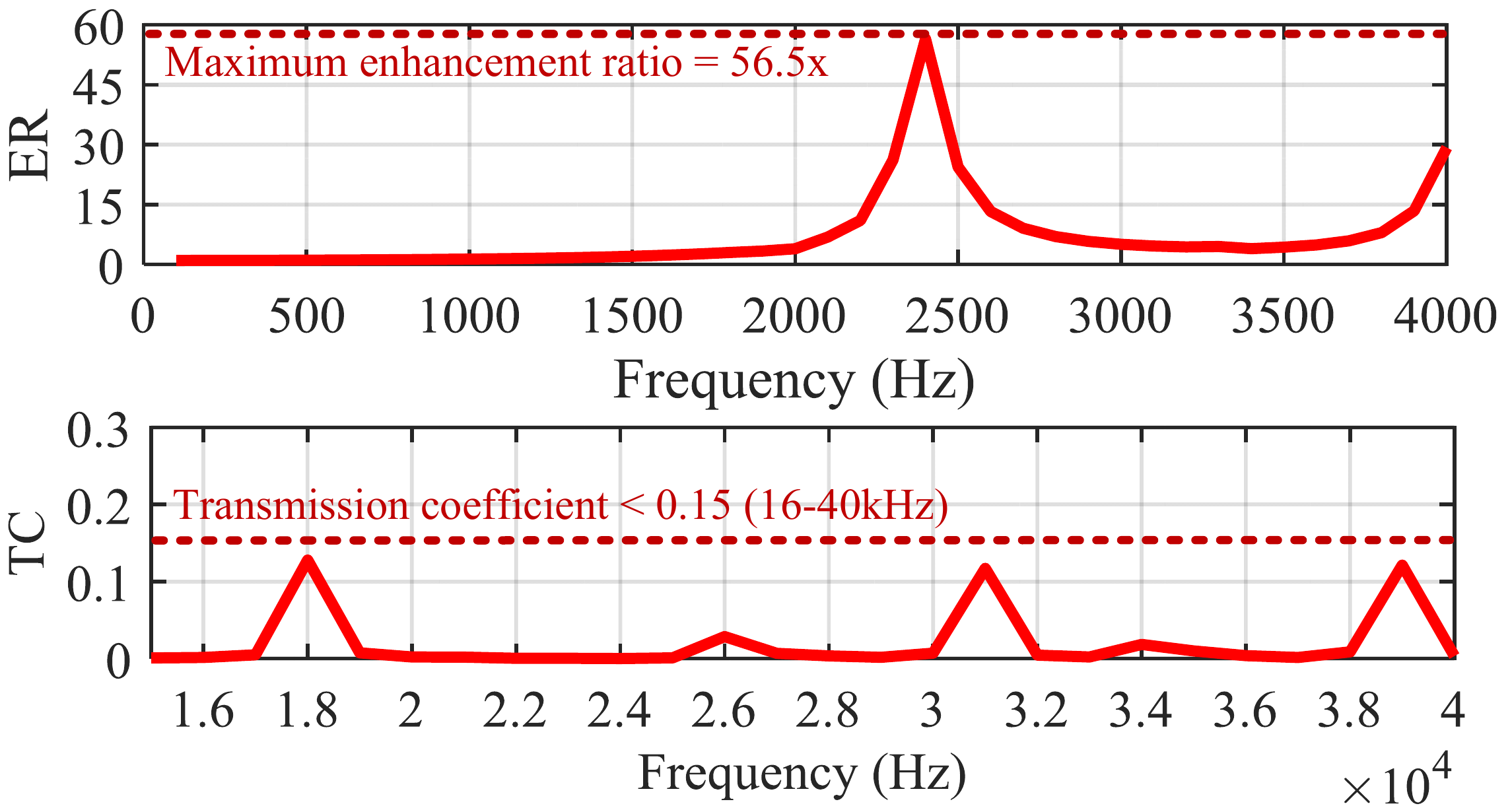}
  \label{meta5A}}
\caption{(a) Smart speaker structure design, (b) its filtering and interference effects.}
\label{LAM-A}
\vspace{-10pt}
\end{figure}

\subsubsection{Universal Structure Design for Mobile Devices} When designing a universal \SystemName structure for mobile devices, we first analyzed their common form factors—typically flat and elongated for easy portability. To preserve this portability, \SystemName adopts a similar shape. In addition, since most mobile devices use a bottom microphone for primary audio capture, the structure must be installed at that location for effective protection.

Figure~\ref{meta44} illustrates our universal framework design. IADM and AADM units are arranged horizontally to fit the device shape. A recessed top secures the device and protects the microphone, while side channels (4 mm × 2 mm) allow legitimate voice signals to pass through. The 5 mm wall blocks 65 dB adversarial signals and resists laser attacks. Core dimensions are $L_1 = 40mm$ , $W_1 = 25 mm$, and $H_1 = 15 mm$. To support different devices, only these three parameters need adjustment. For devices with multiple bottom microphones, additional AADM units can be positioned accordingly to enhance protection.

\cparagraph{\revised{2}{Impact on IADM and AADM performance}} \revised{2}{To evaluate the impact of the \SystemName structure design for mobile devices on defensive effectiveness against IADM and AADM, we used COMSOL to simulate its filtering performance in the ultrasonic range and its interference effects in the low-frequency range. As shown in Figure~\ref{meta44A}, the structure effectively filters inaudible attacks within the 16-40 kHz range. The center frequency of low-frequency enhancement shifted to 2800 Hz, with the gain increasing to 78.8 times. We attribute this change to additional phase shifts along the channel path, which cause constructive interference at specific frequencies~\cite{constructive,constructive2}. This interference shifts the enhanced center frequency and increases the gain. Nevertheless, the variation remains within the acceptable interference frequency range discussed in Section~\ref{Challenge 2}, ensuring that adversarial attacks are effectively disrupted without impairing the recognition of legitimate commands.}

\begin{figure}[!t]
\centering
\subfloat[]{
\includegraphics[scale=0.225]{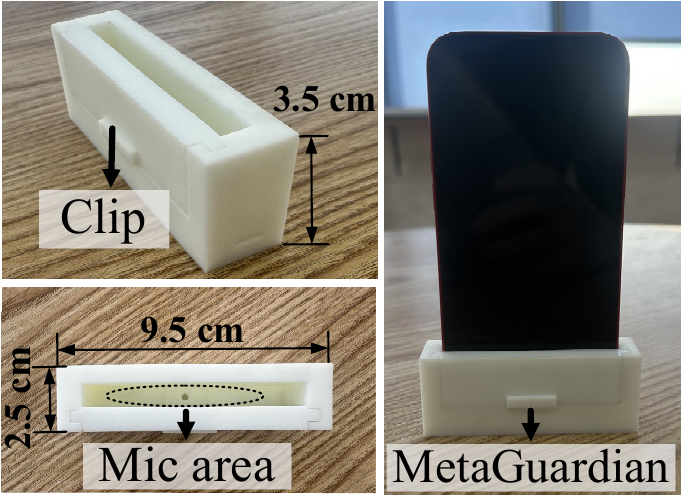}
  \label{prototype1}}
\hfill
\subfloat[]{
		\includegraphics[scale=0.225]{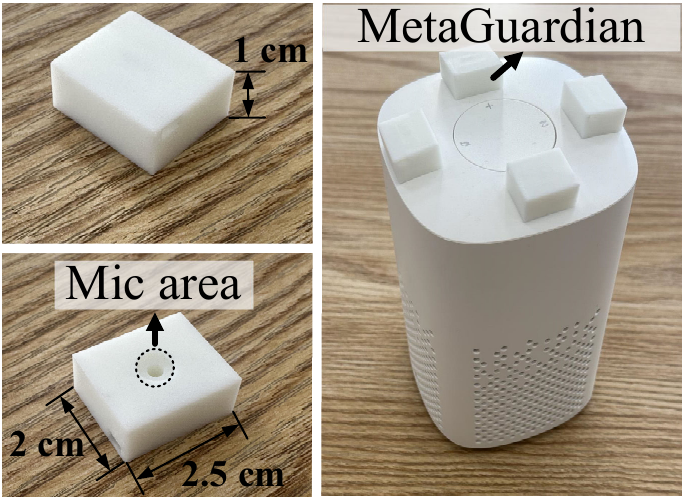}
  \label{prototype2}}
  \hfill
  \subfloat[]{
		\includegraphics[scale=0.225]{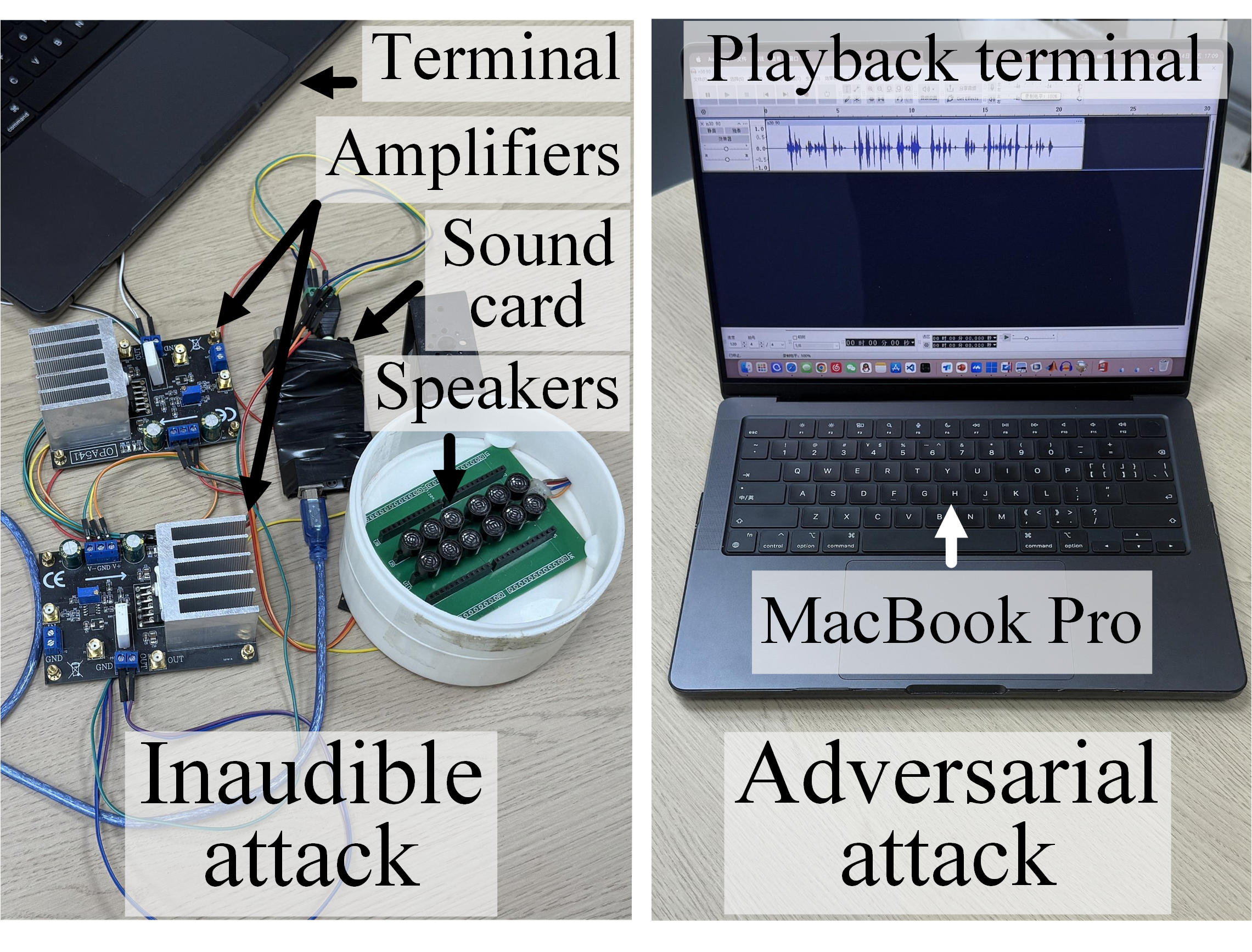}
  \label{prototype3}}
\caption{\revised{3}{\SystemName prototypes for (a) mobile devices, (b) smart speakers, and (c) attack devices.}}

\label{prototype}
 \vspace{-20pt}
\end{figure}

\subsubsection{Universal Structure Design for Smart Speakers}
Smart speakers have microphones concentrated at the top in a circular layout. To fit this design, we developed a compact cubic structure that encloses a single microphone without obstructing buttons. Multiple such units can be combined to protect the entire microphone array.

Figure~\ref{meta5} shows this universal structure. IADM and AADM are arranged in a zigzag pattern to reduce length and avoid blocking buttons. A circular recess at the bottom covers the microphone. The wall thickness matches that of the mobile device structure, allowing attack signals into the internal metamaterial. Dimensions are length $L_2=25mm$, width $W_2=20mm$, height $H_2=10mm$. The circular recess is adjustable to fit microphones of various shapes.

\cparagraph{\revised{2}{Impact on IADM and AADM performance
}} \revised{2}{The COMSOL simulation results for the \SystemName structure design for smart speakers are shown in Figure~\ref{meta5A}. The results confirm that the structure effectively filters inaudible attacks within the 16-40 kHz range. Compared to the \SystemName structure for mobile devices, the center frequency and gain of the low-frequency enhancement show slight variations, likely due to the shorter channel length producing a smaller additional phase shift. These variations are minor and do not affect the overall functionality of the structure.
}

\section{Implementation} \label{chap:7}
\ifx\allfiles\undefined

\else
\fi

The \SystemName prototype is fabricated using resin 3D printing and includes two structural designs tailored for mobile devices and smart speakers (see Figure~\ref{prototype1} and Figure~\ref{prototype2}). The mobile version adopts a slender form to enhance portability, with a front clip to prevent slipping; the smart speaker version is more compact, with a bottom notch to preserve button functionality. Its modular design makes it easy to adapt to different microphone layouts. This structure balances portability and adaptability, and can be extended to various devices by adjusting design parameters.

\revised{3}{Additionally, Figure~\ref{prototype3} shows the devices used in our experiments for inaudible and adversarial attacks: inaudible attacks are amplified through a power amplifier and transmitted via an ultrasonic transducer, while adversarial commands are played through the built-in speaker of a laptop (MacBook Pro).}





\ifx\allfiles\undefined

\else
\fi
\section{Evaluation} \label{chap:6}
\subsection{Experimental Setup and Methodology}

\cparagraph{Test targets} To comprehensively and accurately evaluate the performance of \SystemName in defending against attack signals, we reproduced three types of inaudible attacks with different center frequencies: NUIT~\cite{nuit} (18~kHz), DolphinAttack~\cite{Dolphinlang} (25~kHz), and LipRead~\cite{lipread} (40~kHz), effectively covering the typical attack range of 16-40~kHz. Additionally, we reproduced five authoritative open-source adversarial attacks: ALIF~(2024)\cite{alif}, KENKU~(2023)\cite{kenku}, SMACK~(2023)\cite{smack}, CommanderSong~(2018)~\cite{commandersong} and Devil's Whisper~(2020)~\cite{devil} as well as a laser attack, Light Commands~\cite{lightcommands} (we verified its ability to penetrate \SystemName using a laser pointer and a photosensor). Detailed information on these systems is presented in Table~\ref{targetsystems}. 

\begin{table}[t!]
    \scriptsize
\caption{Tested on nine authoritative attack systems.}
    \label{targetsystems}
    \vspace{1mm}
    \centering
      \setlength{\tabcolsep}{2.9pt} 
    \begin{tabular} {p{2.4cm}p{2cm}p{3.5cm}}
    \toprule      
    \makecell[l]{\textbf{System name}} & \makecell[l]{\textbf{Attack type}} & \makecell[l]{\textbf{Compatible devices}}\\
    \midrule

 \rowcolor{gray!20}\makecell[l]{KENKU~\cite{kenku}} & \makecell[l]{~Adversarial}  & \makecell[l]{iPhone 16 Pro, Pixel 8 Pro,\\ Echo Dot 5th, HomePod mini }\\

\makecell[l]{SMACK~\cite{smack}} &\makecell[l]{~Adversarial}& \makecell[l]{iPhone 14 Pro, Echo Dot 5th}\\

 \rowcolor{gray!20}\makecell[l]{ALIF~\cite{alif}}  & \makecell[l]{~Adversarial} & \makecell[l]{iPhone 14 Pro, Pixel 8 Pro,\\ Echo Dot 5th}\\

\makecell[l]{CommanderSong~\cite{commandersong}}  & \makecell[l]{~Adversarial} & \makecell[l]{iPhone 14 Pro}\\

   \rowcolor{gray!20}\makecell[l]{Devil’s Whisper~\cite{devil}}  & \makecell[l]{~Adversarial} & \makecell[l]{iPhone 16 pro, Pixel 8 Pro\\  Echo Dot 5th, HomePod mini}\\

\makecell[l]{DolphinAttack~\cite{Dolphinlang},\\ NUIT~\cite{nuit}, LipRead~\cite{lipread}} &\makecell[l]{~Inaudible}& \makecell[l]{All devices}\\

 

\rowcolor{gray!20}  \makecell[l]{Light Commands~\cite{lightcommands}} &\makecell[l]{~Laser}& \makecell[l]{All devices}\\
    \bottomrule
    \end{tabular}
\end{table}

\cparagraph{Test devices} To verify the broad applicability of \SystemName, we selected five smartphones and four smart speakers for protection effectiveness testing, covering well-known brands such as Apple~\cite{Apple}, Google~\cite{Googlep}, Xiaomi~\cite{Xiaomi}, Huawei~\cite{Huawei}, and Amazon~\cite{Amazon}. The selected devices include flagship models and highly practical products from these brands in recent years, spanning different types, usage scenarios, and price ranges, and are widely used in personal and home environments. This selection aims to ensure \SystemName's compatibility and effectiveness across various hardware and environments. Detailed specifications of all test devices are listed in Table~\ref{targetsdevices}.

\begin{table}[t!]
    \scriptsize
    \caption{Tested on nine models from six VAs.}
    \label{targetsdevices}
    \vspace{1mm}
    \centering
     \setlength{\tabcolsep}{4.8pt} 
    \begin{tabular} {p{1cm}p{1.6cm}lll}
    \toprule      
    \makecell[l]{\textbf{Manuf.}} & \makecell[l]{\textbf{Model}} & \makecell[l]{\textbf{Type}} & \makecell[l]{\textbf{VA (OS)}}\\
    \midrule
\rowcolor{gray!20}& \makecell[l]{iPhone 16 Pro} &Mobile device & Siri (IOS 18) \\

\rowcolor{gray!20}     & \makecell[l]{iPhone 14 Pro} &Mobile device&  iFlytek (7.0.4062)\\

\rowcolor{gray!20}\makecell[l]{\multirowcell{-3}{~Apple}}  & \makecell[l]{HomePod mini} &Smart speaker& Siri (18.2) \\


    
\makecell[l]{\multirowcell{1}{~Google}}&  \makecell[l]{Pixel 8 Pro}  &Mobile device&  Google Assistant (Android 14)\\

\rowcolor{gray!20}& \makecell[l]{Xiaomi 14}  &Mobile device&  XiaoAI (HyperOS 2)\\ 
    
\rowcolor{gray!20}\makecell[l]{\multirowcell{-2}{~Xiaomi}} &  \makecell[l]{Xiaoai Play 2} &Smart speaker& XiaoAI (1.62.26)\\ 

& \makecell[l]{Mate 60 Pro} &Mobile device& Xiaoyi (HarmonyOS 4)\\

\makecell[l]{\multirowcell{-2}{~Huawei}} & \makecell[l]{AISpeaker 2e} &Smart speaker& Xiaoyi (HarmonyOS 2)\\

\rowcolor{gray!20}\makecell[l]{\multirowcell{1}{~Amazon}} &  \makecell[l]{Echo Dot 5th} &Smart speaker& Alexa (9698496900h)\\

    \bottomrule
    \end{tabular}
\end{table}

\begin{figure}[!t]
\centering
\subfloat[]{
\includegraphics[scale=0.071]{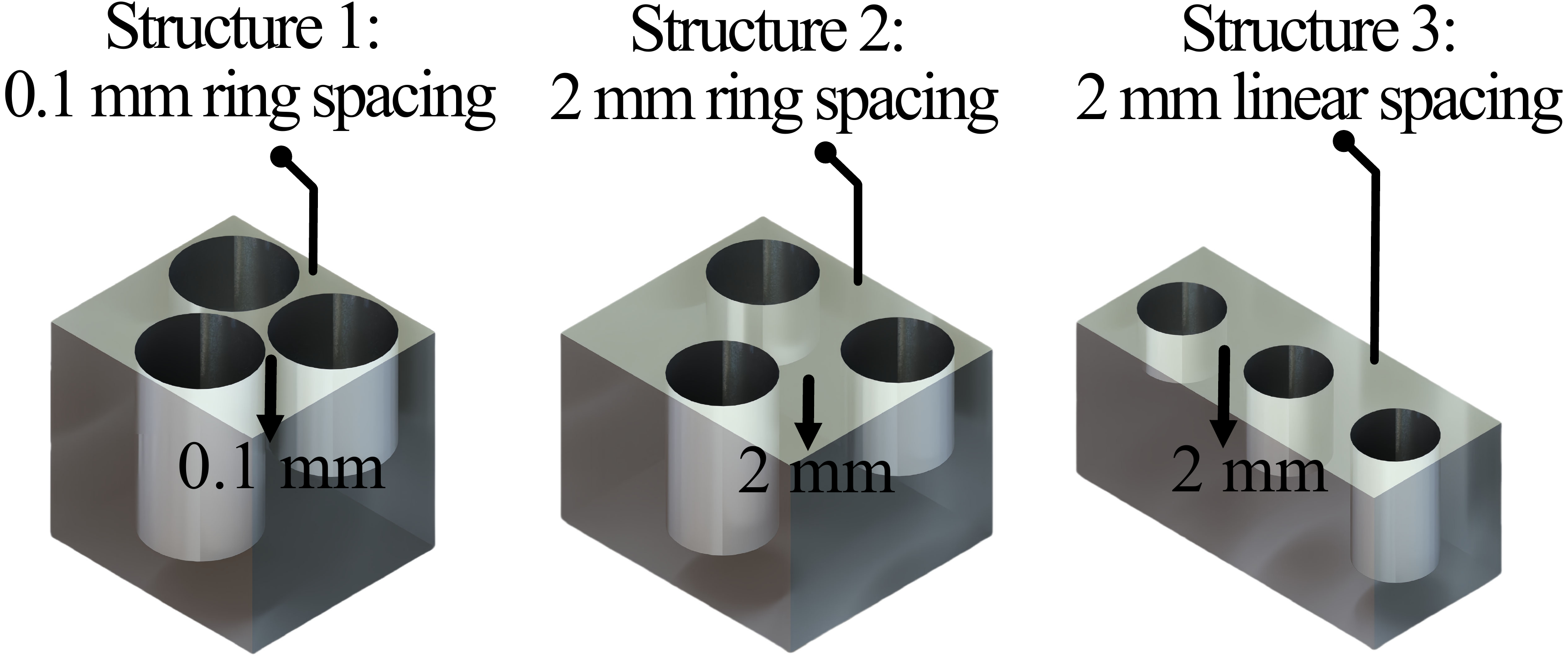}
  \label{RealFliter1}}
\hspace{0.1cm}
\subfloat[]{
		\includegraphics[scale=0.2]{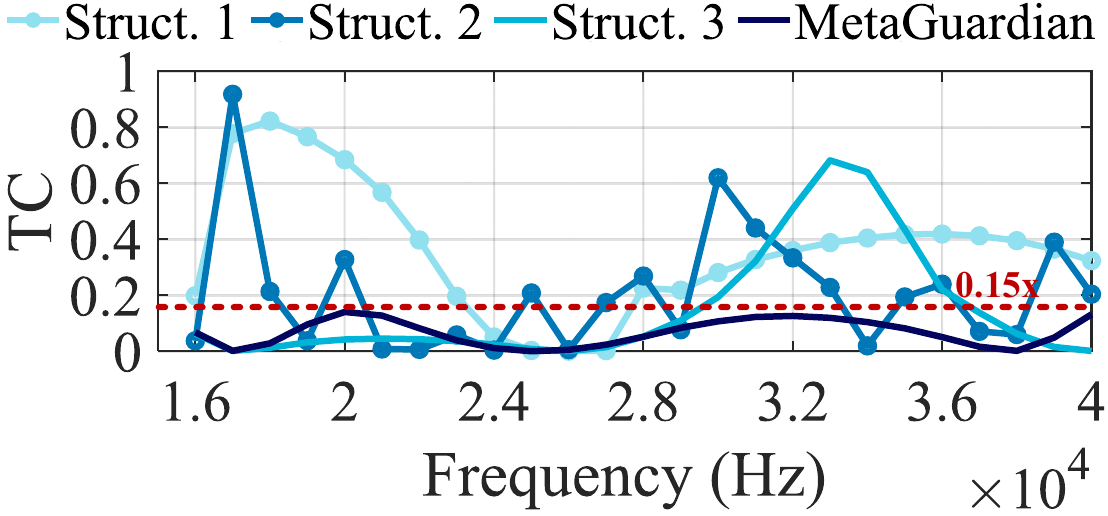}
  \label{RealFliter2}}
\caption{\revised{7}{(a) Three distinct structures, (b) comparison of transmission coefficients (TC).}}
\label{RealFliter}
\vspace{-20pt}
\end{figure}

\cparagraph{Evaluation metrics} We use three distinct evaluation metrics to comprehensively assess the performance of \SystemName across multiple attack scenarios and usage conditions.

\textit{Protection Success Rate (PSR)} measures defense performance against \SystemName by counting failed attacks from 30 attempts with each system.

\textit{Word Interference Rate (WIR)} gauges \SystemName's interference on attack command keywords by the ratio of destroyed to total words. 

\textit{Commands Recognition Rate (CRR)} through evaluates \SystemName's impact on legitimate commands by the ratio of recognized to total commands.




\cparagraph{Experiment design} \revised{3}{We conducted experiments to evaluate the defense capabilities of \SystemName. Adversarial attacks used 65 dB voice commands, including \textit{Open the door, Play music, Make a call, What's the time, Send a message, Turn on the light, Transfer money, Airplane mode on, Navigate to my office,} and \textit{Make a credit card payment.}} \revised{6}{Inaudible attacks transmitted the same commands using a 3-watt ultrasonic speaker. The attack devices are shown in Figure~\ref{prototype3}. Experiments were conducted in an open laboratory with a background noise level of approximately 43 dB to minimize environmental interference and signal loss.}

Table~\ref{experimental setting} details the experiments. Experiments A1 and A2 validated \SystemName's practical applicability. Experiments B1-B3 evaluated its defense against adversarial, inaudible, and laser attacks. Experiments B4-1, B4-2, and B5-1, B5-2 assessed its handling of complex attacks. Experiment C compared \SystemName's advantages to existing defense strategies.

        
        
        


\begin{table*}
    \scriptsize
    \caption{Experimental includes two improvement verifications, seven defense evaluations, and one comparison.}
    \label{experimental setting}
    \vspace{1mm}
    \centering
    \renewcommand{\arraystretch}{1.2}
    \begin{tabular} {p{2.8cm}p{1.5cm}p{3.7cm}p{8.2cm}} \toprule
    \multicolumn{1}{l}{\textbf{Test objectives}}  & \multicolumn{1}{l}{\textbf{Label}} &\multicolumn{1}{l}{\textbf{Test focus}} & \multicolumn{1}{l}{\textbf{Description}}\\
    \midrule
\rowcolor{gray!20}  \makecell[l]{Filtering performance} & A1 (Sec.\ref{A1})& Filtering effect of \SystemName & We tested \SystemName's 16-40 kHz filtering across different unit arrangements.\\ 

\rowcolor{gray!20} \makecell[l]{Impact on normal usage} & A2 (Sec.\ref{A2})& Impact on normal functionality & We measured recognition accuracy during input and playback to assess impact.\\  

& B1 (Sec.\ref{B1}) &  Adversarial attack defense capability & We tested five \SystemName-integrated devices against five adversarial attacks.\\
        
& B2 (Sec.\ref{B2}) & Inaudible attack defense capability & Nine \SystemName-equipped devices were tested against three inaudible attacks.\\

& B3 (Sec.\ref{B3}) & Laser attack defense capability & \SystemName's laser defense was tested using a laser pointer at different angles. \\

& B4 (Sec.~\ref{B4}) & Multi-angle defense capability & We evaluated \SystemName's multi-angle defense against attacks.\\

&  B5 (Sec.~\ref{B5}) &  Precision interference in attacks & We evaluated \SystemName's effectiveness in disrupting command keywords.\\

 \makecell[l]{\multirowcell{-7}{Defense performance}} &  B6 (Sec.~\ref{B6}) &  Anti-interference capability & We tested \SystemName’s defense under environmental interference.\\



\rowcolor{gray!20}   & C1 (Sec.~\ref{C1}) &  Prior work’s reliability affected & To validate our viewpoint, we tested the cross-device reliability of prior work.\\  

\rowcolor{gray!20}  \makecell[l]{\multirowcell{-3}{Compared to prior work}} & C2 (Sec.\ref{C2}) &  Advantages of \SystemName & We compared \SystemName with existing defense strategies in various aspects.\\

\bottomrule
\end{tabular}
\end{table*}
\subsection{System Filtering Performance and Legitimate Signal Impact}

\subsubsection{A1 - Filtering Effect of \SystemName \label{A1}}  

To verify the optimized \SystemName's ability to efficiently filter ultrasonic signals, we used the Avisoft-Bioacoustics CM16/CMPA to measure the ultrasonic signal strength passing through it and compared the effects of different unit arrangements and spacings on filtering (see Figure~\ref{RealFliter1}).

As shown in Figure~\ref{RealFliter2}, \SystemName performs excellently in filtering, consistent with COMSOL simulations. The ring structure with 0.1 mm and 2 mm spacing (Structure~1) is effective only in the 25-30 kHz range, while the linear structure with 0.1 mm spacing (Structure~3) is effective only in the 16-28 kHz range. The results indicate that using a linear arrangement with reduced spacing can significantly enhance the mutual impedance effect.

\subsubsection{A2 - Impact on Normal Usage \label{A2}}  
\revised{2,4}{Before evaluating \SystemName’s defense performance, it is essential to ensure it does not interfere with the input and output of legitimate commands. Therefore, we tested its impact on standard commands such as \textit{What is the weather}, \textit{Play music} and \textit{Open the door}, using voice samples synthesized by Google Cloud TTS~\cite{Googletts} and real speech from 10 male and 10 female volunteers.}

\begin{figure}[t!]
    \begin{minipage}[t]{0.48\linewidth}
        \centering   
        \includegraphics[width=1\linewidth]{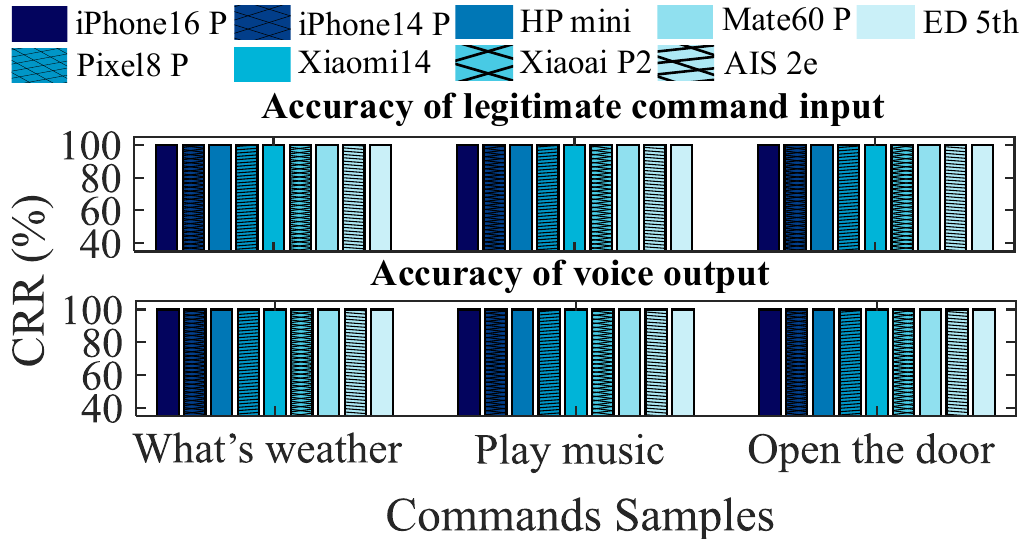}
        \caption{\revised{7}{Impact on commands input \& playback.}}
        \label{legitimate commands}
    \end{minipage}  
    \hfill
    \begin{minipage}[t]{0.48\linewidth}
        \centering        
 \includegraphics[width=1\linewidth]{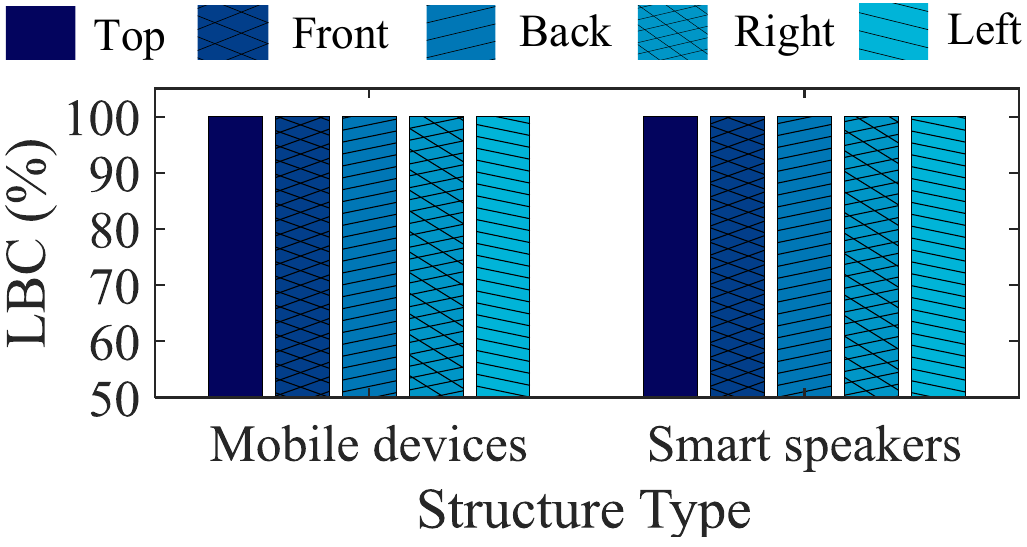}
        \caption{\revised{7}{Laser light-blocking coefficient (LBC)}}
        \label{light blocking coefficients}
    \end{minipage}
     \end{figure}

\revised{2,4}{As shown in Figure~\ref{legitimate commands}, devices equipped with \SystemName successfully responded to all voice commands, and the commands played back were accurately recognized by other devices, achieving a 100\% command recognition rate. These results indicate that devices integrated with \SystemName can operate voice assistant and audio playback functions normally, ensuring a good user experience.}



\subsection{Overall Defense Performance}
\revised{6}{We evaluated the performance of \SystemName against various attacks under controlled conditions. It is important to note that the results of \SystemName were obtained in a controlled environment, which minimizes some variables like user movement and background noise. Performance of \SystemName in unconstrained settings may be influenced by these additional factors.}

\subsubsection{B1 - Adversarial Attack Defense Capability \label{B1}}  
\begin{figure}[t!]
    \centering
    \includegraphics[width=1\linewidth]{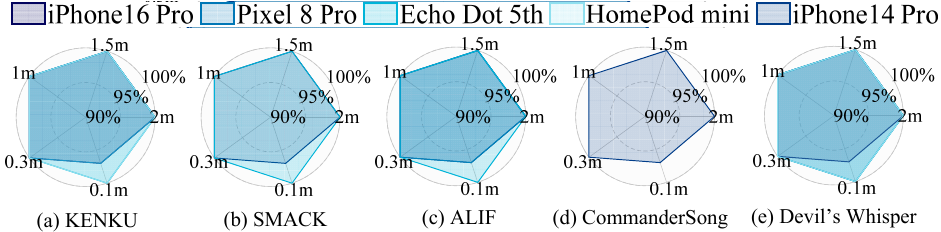}
    \caption{\revised{7}{Adversarial attack defense at various ranges.}}
        \label{ADDefence}
     \vspace{-10pt}
\end{figure}

We evaluated \SystemName against five representative adversarial attacks (Table~\ref{targetsystems}) on five VA-enabled devices. Figure~\ref{ADDefence} presents the attack success rates (PSR) in this setting. \revised{4}{At distances where these attacks typically achieve high success, including KENKU~\cite{kenku} (70\% at 0.3 m), CommanderSong~\cite{commandersong} (82\% at 1.5 m), SMACK~\cite{smack} (64.7\% at 0.5 m), Devil's Whisper~\cite{devil} (90\% at 2 m), and ALIF~\cite{alif} (85.7\% at 0.3 m), \SystemName maintained a 100\% defense success rate. Even under more challenging conditions, with attacks launched from 0.1 m at 65 dB playback volume, defense success remained above 97\% for all five attacks across nine devices.} This robustness is due to the AADM structure's high gain amplification in the 2000--4000 Hz range, which effectively disrupts adversarial signals while preserving the recognition of legitimate commands.

\subsubsection{B2 - Inaudible Attack Defense Capability \label{B2}}  
\begin{figure}[t!]
    \centering
    \includegraphics[width=1\linewidth]{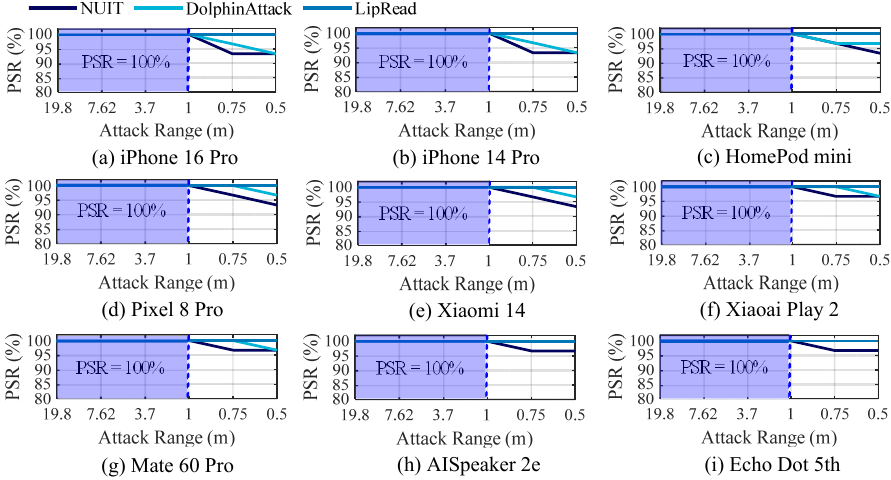}
    \caption{\revised{7}{Inaudible attack defense at various ranges.}}
        \label{InaudibleDefence}
\vspace{-10pt}
\end{figure}

To evaluate \SystemName’s effectiveness against inaudible attacks, we tested nine devices at various distances and recorded the PSR in a controlled environment. The results are shown in Figure~\ref{InaudibleDefence}.
\revised{4}{Within the maximum effective ranges of three common attacks, DolphinAttack achieved 100\% success at 19.8 meters, LipRead 50\% at 7.62 meters, and NUIT over 80\% at 3.8 meters, while \SystemName consistently maintained a 100\% PSR. Even when the attack distance was reduced to 0.5 meters, the PSR for all three attacks remained above 93\%. A slight decline in defense performance at closer distances is attributed to reduced signal attenuation, which allows part of the attack energy to exceed \SystemName's suppression threshold.} However, inaudible attacks typically require conspicuous equipment such as speaker arrays, power amplifiers, and external power supplies, which are difficult to deploy discreetly at short range. As a result, the practical threat in such scenarios remains limited.


\subsubsection{B3 - Laser Attack Defense Capability \label{B3}}  
\begin{figure*}[t!]
\begin{minipage}[t]{0.48\linewidth}
\subfloat[~~Attack angle: 15°]{
\includegraphics[scale=0.226]{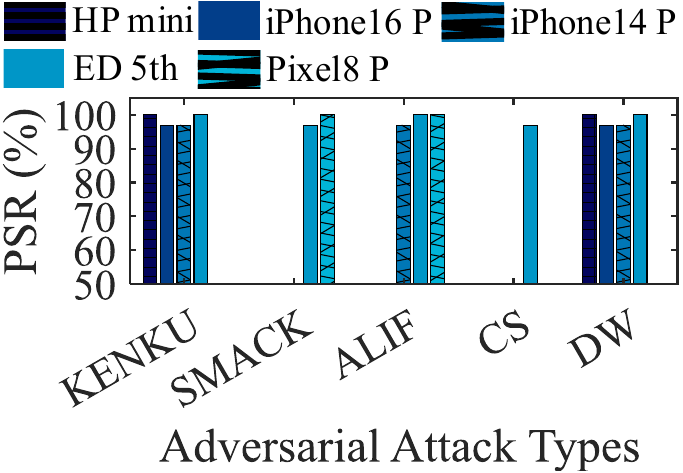}
  \label{Ad15}}
\hspace{0.1mm}
\subfloat[~~Attack angle: 30°]{
		\includegraphics[scale=0.226]{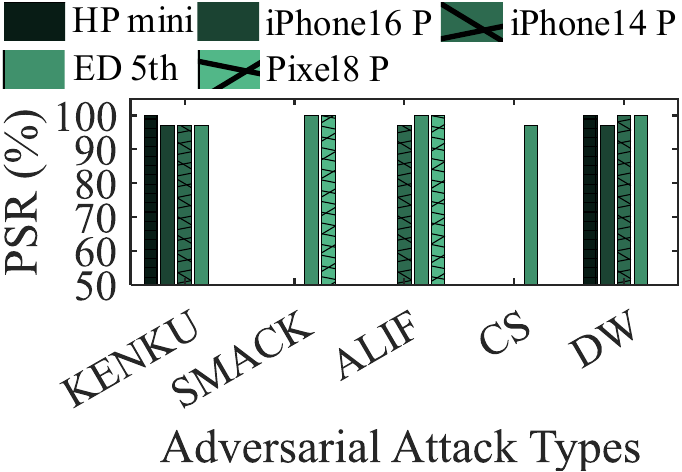}
  \label{Ad30}}
\hspace{0.1mm}
\subfloat[~~Attack angle: 60°]{
		\includegraphics[scale=0.226]{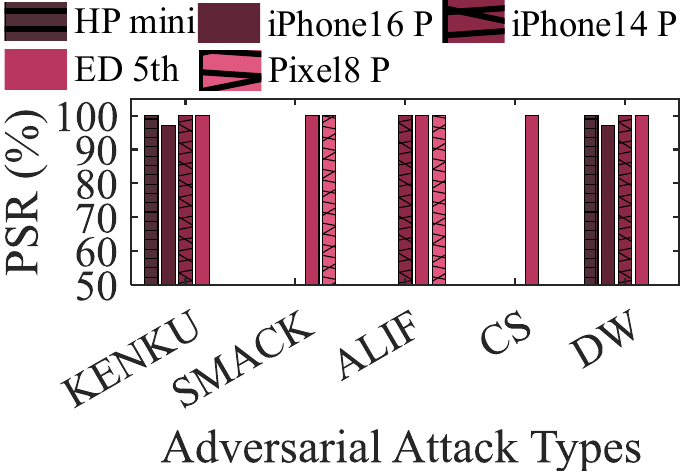}
  \label{Ad60}}
\caption{\revised{7}{Adversarial attack defense at various angles.}}
\label{AdAngle}
\end{minipage}
\hfill
\begin{minipage}[t]{0.48\linewidth}
\subfloat[~~Attack angle: 15°]{
\includegraphics[scale=0.225]{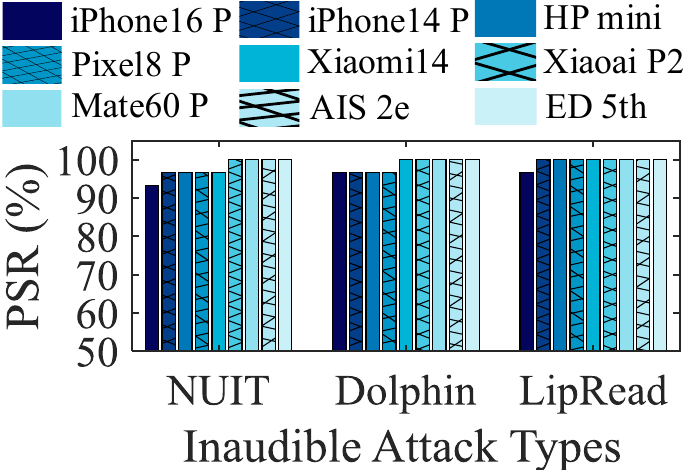}
  \label{Inaudible15}}
\hspace{0.1mm}
\subfloat[~~Attack angle: 30°]{
		\includegraphics[scale=0.225]{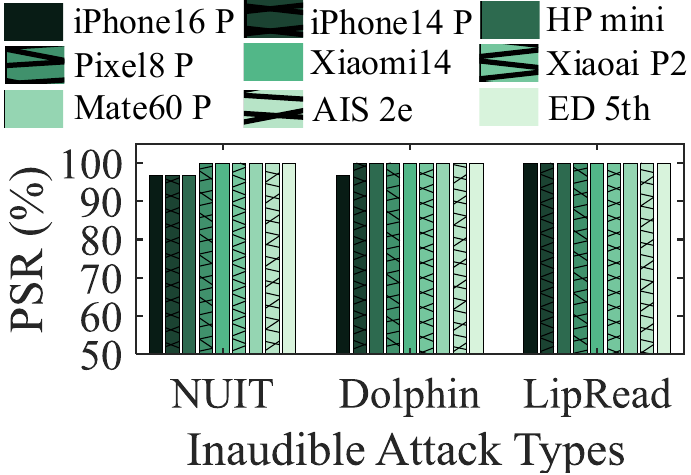}
  \label{Inaudible30}}
\hspace{0.1mm}
\subfloat[~~Attack angle: 60°]{
		\includegraphics[scale=0.225]{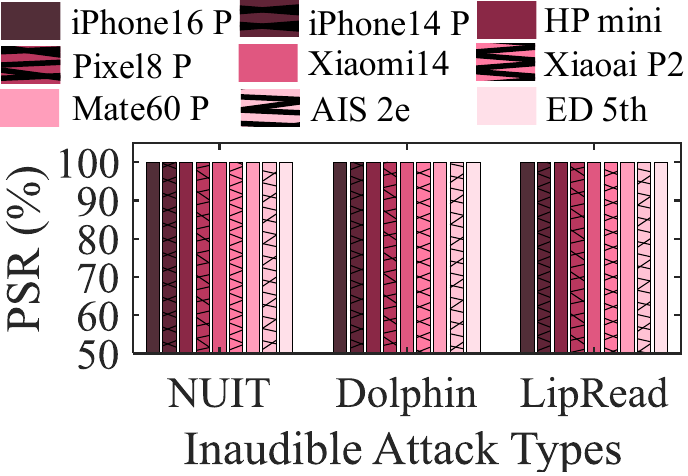}
  \label{Inaudible60}}
\caption{\revised{7}{Inaudible attack defense at various angles.}}
\label{InaudibleAngle}
\end{minipage}

\end{figure*}

\begin{figure}[t!]
    \begin{minipage}[t]{0.48\linewidth}
        \centering   
        \includegraphics[width=1\linewidth]{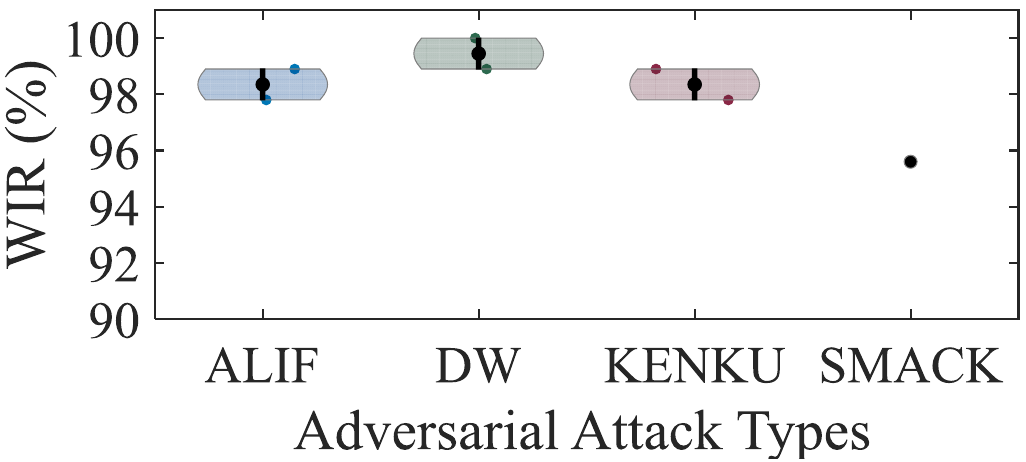}
        \caption{\revised{7}{WIR against adversarial attacks.}}
        \label{Precision1}
    \end{minipage}  
    \hfill
    \begin{minipage}[t]{0.48\linewidth}
        \centering        
 \includegraphics[width=1\linewidth]{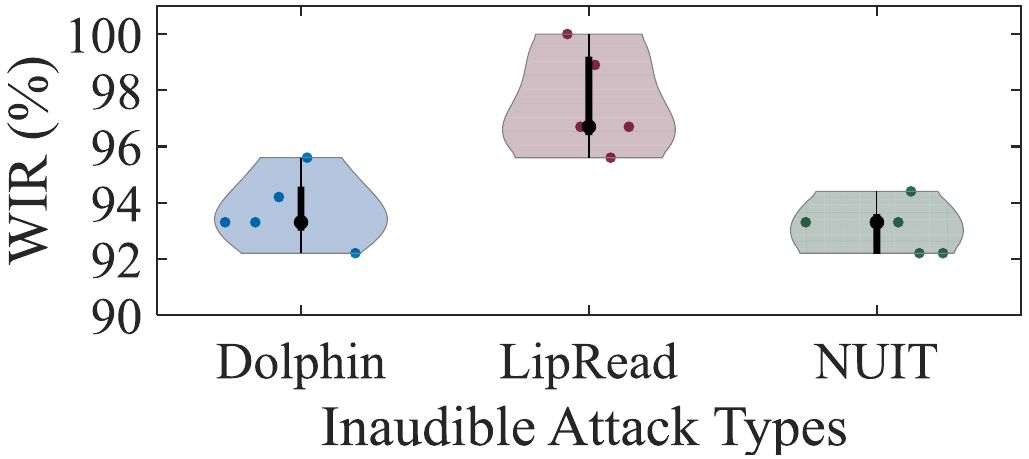}
        \caption{\revised{7}{WIR against inaudible attacks.}}
        \label{Precision2}
    \end{minipage}
     \vspace{-10pt}
     \end{figure}
The main weakness of laser attacks is their inability to penetrate opaque barriers. \revised{6}{We used a 60 mW laser pointer (the same maximum power as in Light Commands~\cite{lightcommands}) to illuminate two \SystemName structures from five angles, and measured their light-blocking coefficients with a TA636A light sensor to evaluate the protective effect.} The results are shown in Figure~\ref{light blocking coefficients}.
At all tested angles, the laser pointer achieved 100\% light blocking when shining on \SystemName, effectively preventing laser transmission. Analysis shows that \SystemName significantly attenuates the laser energy through optical absorption and refraction, blocking the attack commands carried by the laser and causing the attack to fail.

\subsubsection{B4 - Multi-angle Defense in Adversarial and Inaudible Attacks \label{B4}}  
In real-world scenarios, attacks may come from multiple directions. To evaluate \SystemName’s defense performance at different angles, we conducted adversarial and inaudible attacks from 15°, 30°, and 60° angles at distances of 0.1 m and 0.5 m, respectively, and recorded the attack success rate (PSR). The results are shown in Figures~\ref{AdAngle} and \ref{InaudibleAngle}.
For adversarial attacks, \SystemName consistently achieved a PSR exceeding 96\% across all tested angles. For inaudible attacks, the PSR remained above 93\% at all angles, with defense effectiveness improving as the angle increased, reaching 100\% at 60°. This improvement is attributed to the optimized wall thickness design in \SystemName (see Section~\ref{Challenge 3}), which effectively blocks some attack signals, forcing the remaining signals to pass through the metamaterial’s internal structure where they encounter interference.

\subsubsection{B5 - Precision Interference in Adversarial and Inaudible Attacks \label{B5}}  

When defending against multi-keyword attacks, precise interference with each keyword is essential. To evaluate \SystemName’s effectiveness, we launched adversarial and inaudible attacks at 0.1 m and 0.5 m on various speech-to-text mobile devices (iPhone 16 Pro, iPhone 14 Pro, Pixel 8 Pro, Xiaomi 14, Mate 60 Pro) and measured the WER, as shown in Figures~\ref{Precision1} and~\ref{Precision2}.
\SystemName achieves a WER over 95\% in adversarial attacks (deviation $\leq 5\%$) by dispersing and absorbing keyword signal energy to hinder recognition. In inaudible attacks, it maintains WER above 92.5\% (deviation $\leq 7\%$), demonstrating stable, effective defense against complex attacks.

\subsubsection{B6 - Anti-interference Capability \label{B6}}  

\begin{figure}[t!]
    \begin{minipage}[t]{0.48\linewidth}
        \centering   
        \includegraphics[width=1\linewidth]{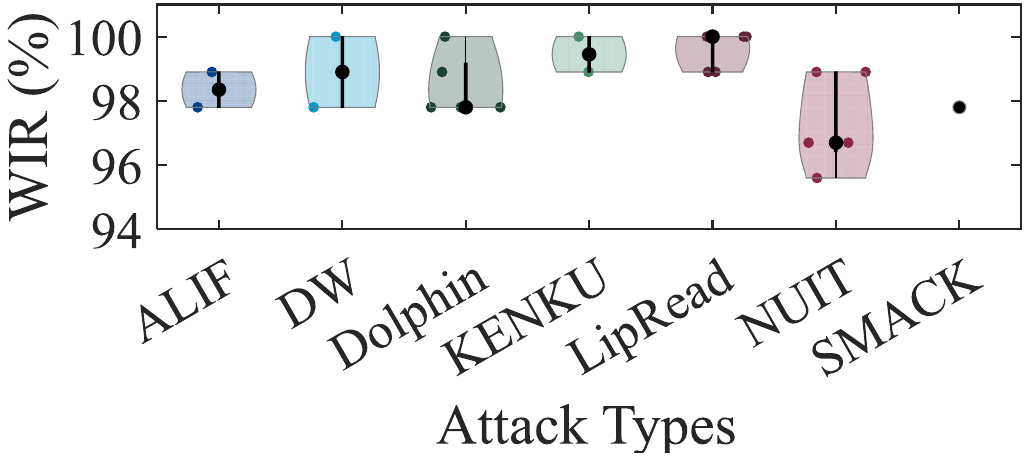}
        \caption{\revised{7}{WIR in a noisy environment.}}
        \label{noise1}
    \end{minipage}  
    \hfill
    \begin{minipage}[t]{0.48\linewidth}
        \centering        
 \includegraphics[width=1\linewidth]{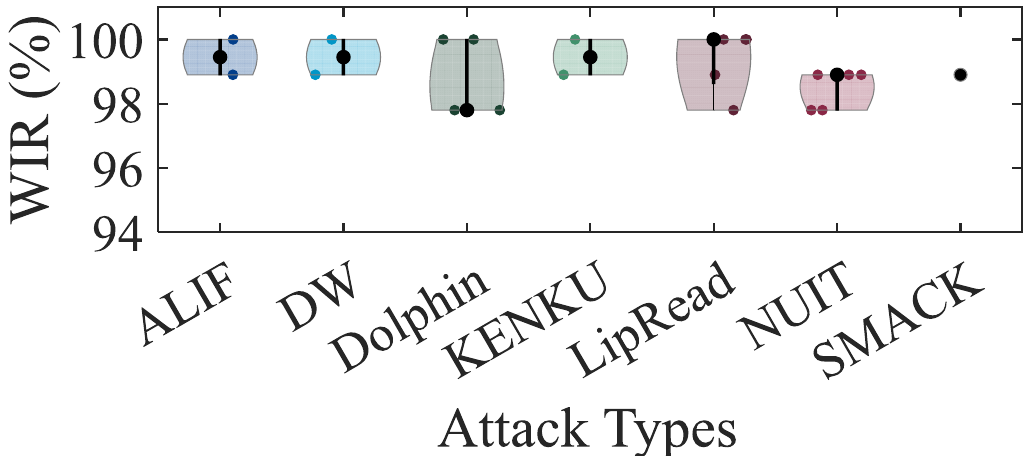}
        \caption{\revised{7}{WIR in a mobile environment.}}
        \label{move2}
    \end{minipage}
     \vspace{-10pt}
     \end{figure}

\revised{3}{To evaluate the system's anti-interference capability in outdoor conditions, we conducted tests in an environment with approximately 75~dB ambient noise. Volunteers carrying devices equipped with \SystemName moved at a speed of 2~m/s while attacks were launched, and the word identification rate (WIR), a higher-is-better metric, was measured, as shown in Figures~\ref{noise1} and~\ref{move2}. \SystemName achieved a WIR of 98\% against adversarial attacks and over 95\% against inaudible attacks in noisy conditions. While in motion, the WIR for both attack types exceeded 97\%, demonstrating high reliability. 
\SystemName employs a passive structure that requires no signal analysis. By altering the phase of sound waves through its material properties, it nonselectively interferes with specific frequencies. This approach is inherently resistant to variations in noise, temperature, and other environmental factors, enabling stable and continuous protection.
}



\subsection{Compared to Prior Work \label{C}}
\subsubsection{C1 - Reliability Impacted by Microphone Differences of Prior Work \label{C1}}  
\begin{figure}[!t]
\centering
\subfloat[]{
\includegraphics[scale=0.215]{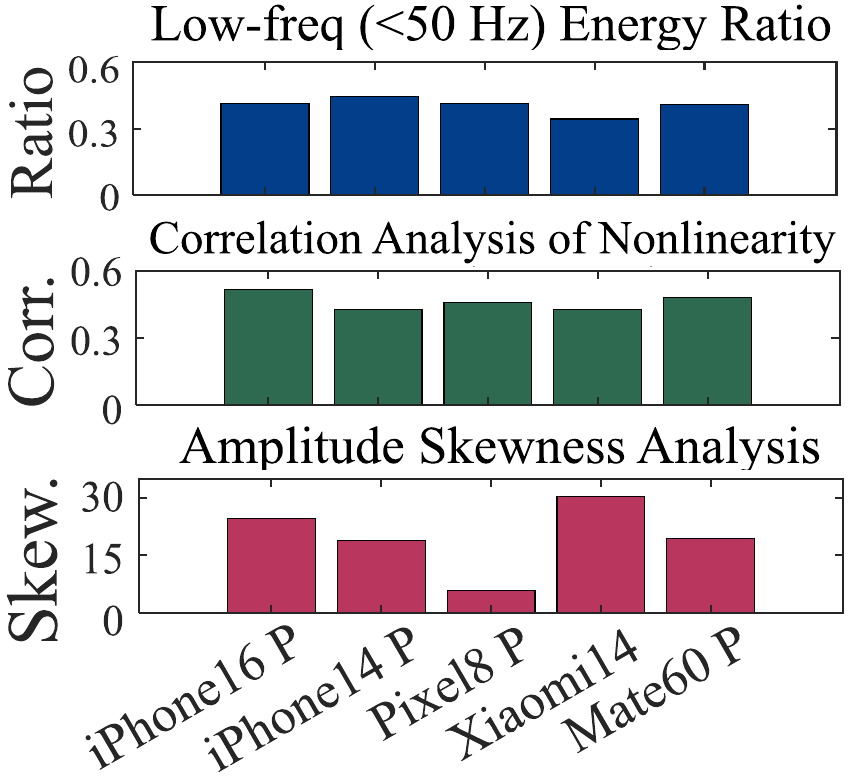}
  \label{C11}}
\hspace{0.1cm}
\subfloat[]{
		\includegraphics[scale=0.35]{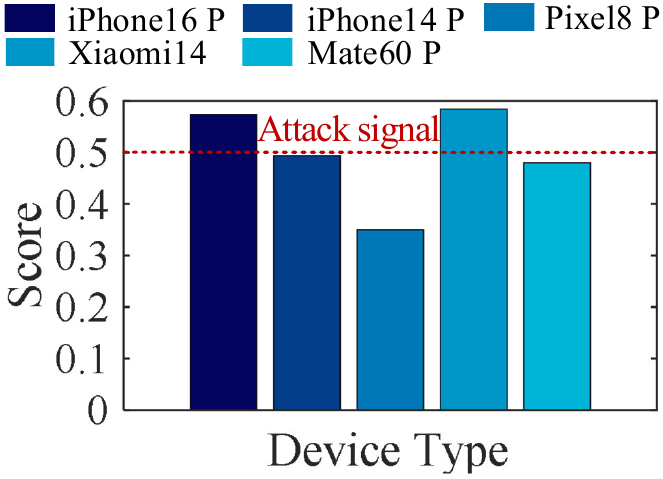}
  \label{C12}}
\caption{\revised{7}{(a) Feature responses of different devices to the same command, (b)comprehensive scores.}}
\label{Comprehensive score}
\vspace{-10pt}
\end{figure}

Variations in the frequency response of microphones across different devices cause significant differences in the received audio signals, affecting the accuracy of defense methods based on signal feature detection~\cite{Dolphinlang, eararray, vocalprint, normaldetect, lipread}. We selected the classic LipRead method~\cite{lipread} for testing (other defense methods use similar signal feature extraction approaches). Under the same environment, the “turn on hotspot” command was recorded 30 times using different devices, and the average values of three features—power, autocorrelation coefficient, and amplitude skew—were calculated and combined into a comprehensive score. The results show that the differences in these three features across devices reached 17\%, 22\%, and 80.97\%, respectively, causing some devices (such as iPhone 14 Pro, Xiaomi 14, and Pixel 8 Pro) to misclassify the attack command as legitimate (see Figure~\ref{C12}). In contrast, \SystemName defends the microphone directly with a physical structure, avoiding the impact caused by hardware differences.

\subsubsection{C2 - Advantages of \SystemName \label{C2}}  
We conducted a comparative analysis of \SystemName and recent defense approaches to evaluate its advantages in usability. As shown in Table~\ref{COMPARE}, five mainstream software-based defenses require disabling the voice assistant upon detecting an attack, which disrupts normal usage and is difficult to deploy in closed systems. \revised{4}{Although these methods achieve over 90\% defense success rates, they are, as discussed in Section~\ref{C1}, susceptible to variations in microphone characteristics across devices. In contrast, \SystemName employs a passive physical structure that directly disrupts attack signals outside the microphone, without modifying system logic or relying on software support, offering greater stability and broader compatibility.}

\revised{7}{Existing hardware-based defense methods, such as AIC~\cite{AIC}, VocalPrint~\cite{vocalprint}, and the approach proposed by Sahidullah et al.~\cite{Sahidullah}, achieve defense success rates above 90\%.} \revised{4}{However, they rely on active components such as speaker arrays, millimeter-wave radar, or continuously worn headsets, which reduce system reliability and portability. In contrast, \SystemName adopts a passive design that requires no device modifications or user intervention, offering strong compatibility and adaptability. Moreover, \SystemName can be seamlessly integrated with existing software and hardware defenses, demonstrating excellent synergy across different defense strategies.}

\section{Discussion and Future Work} \label{chap:9}

\cparagraph{\revised{5}{Mismatch with dynamic attacks}} \revised{5}{Current system have fixed filtering or amplification bands, making them ineffective against adaptive attacks like frequency hopping, and exposing critical vulnerabilities against complex threats.}

\cparagraph{\revised{4,6}{Portability limitations}} \revised{4,6}{While these designs offer some protection, they are often bulky, heavy, and unattractive, reducing portability and user experience. They are unsuitable for scenarios requiring compactness and discretion.}

\cparagraph{\revised{5}{Electromagnetic interference (EMI) defense}} \revised{5}{The current design remains vulnerable to EMI attacks, which can inject malicious signals without using the acoustic channel, weakening the effectiveness of \SystemName and similar acoustic-based defenses.}

\cparagraph{\revised{4}{Impact on ultrasonic sensing}} \revised{4}{Filtering out the 16–40 kHz band may interfere with ultrasonic sensing functions in modern voice assistants, such as proximity detection, gesture recognition, and acoustic analysis, thereby affecting user experience with these features.}

\cparagraph{Future directions} To address current limitations, future work can explore tunable acoustic metamaterials using piezoelectric materials or shape memory alloys, enabling real-time, electrically controlled impedance adjustment to balance defense and sensing, and adapt to dynamic or frequency-hopping attacks.  To resist EMI attacks, shielding or active suppression can be incorporated to build a multilayer defense for enhanced practicality and security.





\begin{table}[t!]
    \scriptsize
    \caption{\revised{7}{Performance compared to prior research}}
    \label{COMPARE}
    \vspace{1mm}
    \centering
    \setlength{\tabcolsep}{2pt} 
    \begin{tabular} {llllll}
    \toprule
    \textbf{\makecell[l]{System\\ name}} & 
    \textbf{\makecell[l]{Function\\ intact}}  & 
    \textbf{\makecell[l]{Closed\\ system def.}} &  
    \textbf{\makecell[l]{No\\ modify}} & 
    \textbf{\makecell[l]{Portable}}   &
     \textbf{\makecell[l]{Multi-attack\\ def.}}

    \\  
    \midrule
    \rowcolor{gray!20}   DolphinAttack~\cite{r8}&  No & No  & Yes& Yes& No\\
    \rowcolor{gray!20}   LipRead~\cite{lipread}&  No & No  & Yes& Yes& No\\
    \rowcolor{gray!20}   NormDetect~\cite{normaldetect}&  No & No & Yes& Yes& No \\
    \rowcolor{gray!20}  EarArray~\cite{eararray} & No & No & Yes& Yes & No\\
    \rowcolor{gray!20}  VoShield~\cite{voshield} & No & No  & Yes& Yes& Yes\\

    AIC \cite{AIC} & Yes & No  & Yes & No & No \\
    VocalPrint \cite{vocalprint} & Yes & Yes  & Yes & No& Yes\\
    Sahidullah et al. \cite{Sahidullah} & Yes & Yes  & Yes & No& Yes\\

    \rowcolor{gray!20} \textbf{\SystemName}& \textbf{Yes}& \textbf{Yes}& \textbf{Yes}&\textbf{Yes} & \textbf{Yes}\\  
    \bottomrule
    \end{tabular}
\end{table}

\section{Conclusion} \label{chap:12}
We have presented \SystemName, an acoustic metamaterial-based VA protection system that blocks inaudible, adversarial, and laser attacks without software support. By leveraging mutual impedance, it expands the filtering range and reduces size, enabling frequency-targeted defense while maintaining audio transmission. Its adaptable design supports diverse devices. Experiments confirm \SystemName is effectively in protecting VA systems across attack types and hardware platforms, making it a reliable, practical solution.

\bibliographystyle{ACM-Reference-Format}
\bibliography{references}

\end{document}